# Lattice-Boltzmann-Driven Physics-Informed Neural Networks for Droplet Wettability on Rough Surfaces

**Ganesh Sahadeo Meshram[1], Partha P Chakrabarti[2], Suman Chakraborty[1]***

[1]Department of Mechanical Engineering, IIT Kharagpur, Kharagpur, 721302, India

[2]Department of Computer Science & Engineering, IIT Kharagpur, Kharagpur, 721302, India

Corresponding Author: suman@mech.iitkgp.ac.in

**Abstract**

We introduce a Lattice-Boltzmann-Driven Kinetic Physics-Informed Neural Network (K-PINN) for predictive modeling of droplet dynamics on rough and structured surfaces, directly embedding the discrete Boltzmann-BGK equation into the learning architecture. Departing fundamentally from conventional PINNs constrained by macroscopic continuum equations, the framework operates at the mesoscopic kinetic level, preserving essential Lattice-Boltzmann physics, including Shan–Chen pseudopotential interactions, within a data-efficient neural surrogate. Across randomly rough substrates and periodically textured pillar arrays, the K-PINN accurately captures nontrivial interfacial phenomena such as contact-line pinning, anisotropic spreading, and capillary hysteresis, while maintaining strict physical consistency, including mass conservation within 1.5%. A U-Net–based encoder–decoder architecture yields 50–75% error reduction relative to baseline networks, achieving near-perfect agreement with high-resolution Lattice-Boltzmann simulations ($L_2 \approx 0.021$–$0.026$, $R^2 \approx 0.999$). Robust convergence over diverse surface morphologies is ensured through curriculum-guided training and adaptive two-phase optimization. Once trained, the model delivers real-time predictions exceeding $10^4$ evaluations per second, offering orders-of-magnitude acceleration over direct numerical simulation and enabling rapid parametric exploration and inverse surface design. By unifying kinetic theory and physics-informed learning, this work establishes a new paradigm for fast, physically faithful modeling of multiphase flows on complex surfaces.

## 1. Introduction

Predicting multiphase droplet dynamics on rough and structured surfaces remains one of the most challenging problems in interfacial fluid mechanics, with direct relevance to energy systems, advanced coatings, biomedical diagnostics, and micro-nano fluidic technologies[1–4]. At its core, the problem is governed by a tightly coupled hierarchy of phenomena: surface morphology modulates local wettability, wettability controls contact-line motion, and contact-line dynamics dictates macroscopic spreading, pinning, and hysteresis[5–7]. These interactions span molecular, mesoscopic, and continuum scales, rendering purely macroscopic descriptions fundamentally incomplete.

Classical continuum approaches based on the Navier–Stokes equations, even when augmented with sophisticated contact-angle models or phase-field formulations, encounter intrinsic limitations in resolving contact-line pinning, roughness-induced hysteresis, and phase separation on complex substrates[8,9]. In contrast, the Lattice Boltzmann Method (LBM) has emerged as the method of choice for multiphase droplet–surface interactions precisely because it operates at a mesoscopic kinetic level, where interfacial forces, wettability effects, and complex geometries are handled naturally through discrete distribution functions and interaction potentials such as the Shan–Chen pseudopotential[10–13]. This mesoscopic fidelity, however, comes at a steep computational cost. High-resolution LBM simulations scale poorly with geometric complexity, parameter dimensionality, and three-dimensionality, making exhaustive parametric studies, inverse design, and real-time prediction impractical[14,15].

The above conflict between physical fidelity and computational tractability has motivated the rapid adoption of machine learning–based surrogates in fluid mechanics[16–21]. Early efforts relied on convolutional and recurrent neural networks to emulate flow evolution from large volumes of simulation data. While successful in accelerating prediction, such models remain fundamentally data-driven, often fragile outside training regimes, and largely agnostic to the governing physics that underpin multiphase behavior. The emergence of physics-informed neural networks (PINNs) marked a decisive step forward by embedding governing equations directly into learning architectures, dramatically reducing data requirements and improving extrapolative robustness[22–24]. PINNs have since been applied successfully to two-phase Navier–Stokes systems, phase-field models, and reactive transport problems[25–29].

Despite significant advancements in these fronts, a critical limitation persists: nearly all existing PINNs enforce physics exclusively at the macroscopic continuum level. As a result, they inherit the same conceptual blind spots as continuum solvers, struggling to capture the mesoscopic mechanisms, namely, distribution-function dynamics, discrete forcing, and kinetic relaxation, which fundamentally govern interfacial phenomena in LBM-based multiphase flows[30,31]. In parallel, surrogate LBM models trained on simulation data bypass governing equations altogether, sacrificing physical consistency for speed and exhibiting limited generalizability across surface morphologies and wettability regimes[32–35].

Recent hybrid efforts sit uneasily between these extremes. Data-driven surrogates trained on LBM outputs reproduce interface shapes but lack rigorous physical regularization, while continuum-level PINNs successfully replicate droplet spreading only on idealized smooth surfaces[36]. Notably absent from the literature is a framework that treats the Lattice Boltzmann equation itself not its macroscopic limit as the governing physics to be learned. To date, no PINN formulation explicitly embeds the discrete Boltzmann–BGK equation with mesoscopic forcing into the neural architecture, nor targets rough and structured surfaces where kinetic effects dominate contact-line dynamics, resulting in a foundational disconnect. This deficit stems from the fact that multiphase wetting on rough substrates is governed not merely by continuum balance laws but by kinetic relaxation, nonlocal interparticle forcing, and discrete velocity effects all intrinsic to LBM and invisible to macroscopic PINNs. The absence of a learning framework capable of enforcing these principles constitutes a central bottleneck in advancing fast yet faithful predictive models for droplet–surface interactions[37–39].

Here, we address this gap by introducing a Lattice-Boltzmann-Driven Kinetic Physics-Informed Neural Network (K-PINN). Unlike conventional PINNs or LBM surrogates, the framework embeds the discrete Boltzmann–BGK equation with Shan–Chen pseudopotential forcing directly into the loss function, enforcing physics at the distribution-function level. The neural network predicts mesoscopic particle distribution functions at arbitrary spatiotemporal locations, from which macroscopic density and velocity fields are reconstructed via moment summation, ensuring simultaneous kinetic and continuum consistency. Sparse, strategically sampled LBM data are used only to anchor learning, not to replace governing physics.

We demonstrate the capability of the K-PINN framework on droplet dynamics over randomly rough and periodically textured surfaces, capturing contact-line pinning, anisotropic spreading, and capillary hysteresis with high fidelity while achieving orders-of-magnitude speedup over direct LBM simulations. By unifying kinetic theory, physics-informed learning, and deep neural approximation, this work establishes a new computational paradigm that is neither a surrogate LBM nor a conventional PINN, but a physically grounded kinetic emulator for complex multiphase wetting phenomena.

## 2. Methods

### 2.1 Why Kinetic PINNs?

PINNs have fundamentally altered how governing laws are embedded into learning frameworks. By enforcing conservation laws and constitutive relations through loss functions, PINNs offer a compelling alternative to purely data-driven surrogates[40]. However, their formulation has remained almost exclusively anchored to macroscopic continuum equations, most commonly the Navier–Stokes system augmented by phase-field or sharp-interface models[41–43]. While sufficient for smooth interfaces and weak wetting heterogeneity, this paradigm becomes intrinsically inadequate when droplet dynamics are controlled by microscale roughness, contact-line pinning, and nonlocal interfacial forces.

At the heart of the above limitation lies a scale mismatch. Macroscopic PINNs enforce averaged balance laws that implicitly assume local thermodynamic equilibrium and smooth constitutive closures. In contrast, droplet–surface interactions on rough and structured substrates are governed by mesoscopic kinetic processes: discrete momentum exchange at the interface, nonlocal interparticle interactions, and relaxation toward equilibrium over finite lattice times. These effects manifest most strongly near contact lines, within confined geometries, and across wettability transitions precisely the regimes where continuum PINNs exhibit degraded accuracy or require ad hoc corrections[44].

The LBM resolves this gap by operating at the kinetic level, evolving particle distribution functions rather than macroscopic fields[45]. Interfacial tension, phase segregation, and wettability emerge naturally through collision–streaming dynamics and forcing schemes such as the Shan–Chen pseudopotential, without explicit interface tracking or empirical contact-line models. Crucially, these mechanisms are not reducible to closed-form macroscopic equations without loss of fidelity. Consequently, enforcing only the macroscopic limits of LBM within a PINN framework discards the very physics that makes LBM successful for multiphase wetting.

Kinetic PINNs address this deficiency by elevating the discrete Boltzmann equation itself to the status of the governing constraint. Instead of learning velocity and pressure fields subject to continuum residuals, a kinetic PINN learns the evolution of particle distribution functions constrained by the Boltzmann–BGK equation with explicit forcing. Macroscopic observables are recovered as moments of these distributions, ensuring that continuum consistency is a consequence of kinetic fidelity not an imposed approximation. This inversion of hierarchy fundamentally distinguishes kinetic PINNs from both traditional PINNs and LBM-trained surrogates.

From a learning perspective, kinetic enforcement offers several decisive advantages. First, it provides stronger inductive bias, constraining the hypothesis space far more tightly than macroscopic equations and significantly improving generalization across surface morphologies and wettability regimes[46–49]. Second, it embeds mass and momentum conservation at the distribution-function level, eliminating the need for auxiliary regularization terms. Third, it naturally accommodates complex geometries and boundary conditions, as these enter through streaming and forcing rather than explicit geometric parametrization. Most importantly, kinetic PINNs preserve the emergent nature of interfacial phenomena. Contact-line pinning, capillary hysteresis, and anisotropic spreading arise organically from the learned kinetic dynamics, rather than being prescribed through empirical boundary laws. This makes kinetic PINNs uniquely suited for rough and structured surfaces, where wettability is not a material constant but an emergent outcome of multiscale interactions.

In this sense, kinetic PINNs are not a refinement of continuum PINNs, nor a compressed surrogate of LBM. They constitute a distinct modeling paradigm in which learning occurs at the same physical level as the most accurate numerical methods, while retaining the speed and flexibility of neural approximators. By embedding mesoscopic physics directly into the learning architecture, kinetic PINNs bridge the longstanding divide between high-fidelity multiphase simulation and rapid predictive modelling, thereby unlocking new possibilities for parametric exploration, inverse design, and real-time decision-making in complex wetting systems.

### 2.2 Problem definition and computational domain

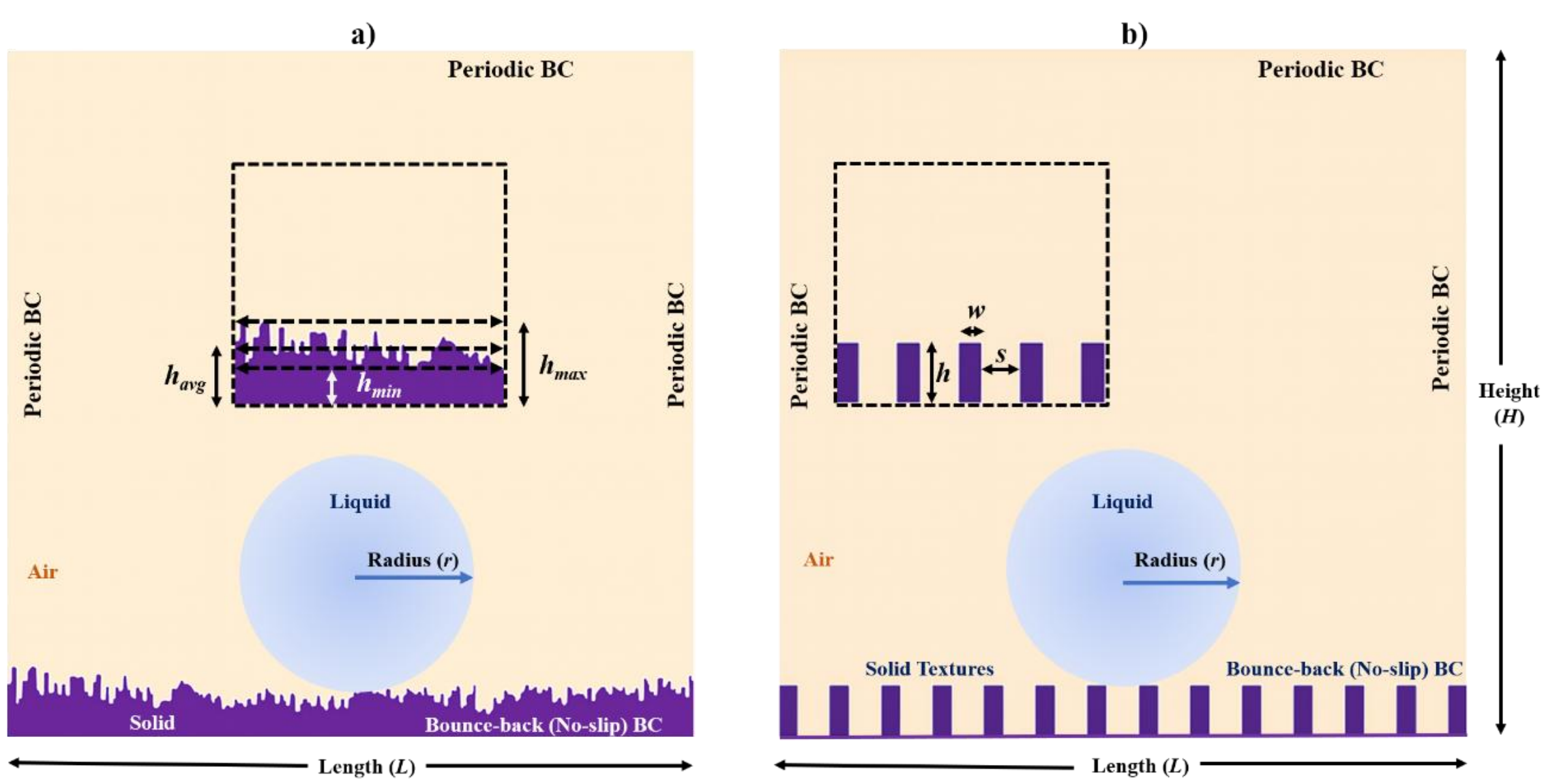

**Figure 1.** Schematic of computational domain showing (a) stochastic rough surface and (b) periodic textured surface configurations with boundary conditions.

To investigate droplet dynamics on textured surfaces, we consider a two-dimensional domain terms of lattice units $\Omega = \{(x, y) \mid x \in [0, L], y \in [0, H]\}$, with domain dimensions $L = 400\Delta x, H = 200\Delta x$, where $\Delta x$ is the lattice spacing. Time is discretized as $\Delta t = 1$, and all quantities are expressed in lattice units unless stated otherwise. The flow is modeled using the D2Q9 lattice, where the discrete velocity set is

$$c_i = \begin{cases} (0,0), & i=0 \\ (\pm 1,0),(0,\pm 1), & i=1-4 \\ (\pm 1,\pm 1), & i=5-8 \end{cases} \tag{1}$$

The computational domain contains three phases, liquid phase (droplet) with density $\rho_l = 6.5$, gas phase (air) with density $\rho_g = 0.38$, solid phase representing either stochastic roughness or periodic textures. The density ratio $\rho_l/\rho_g \approx 17$ allows stable interface capturing while maintaining numerical robustness in the Shan-Chen framework[45,50,51]. The lattice Boltzmann equation with BGK collision operator with evolution of the particle distribution function $f_i(\mathbf{x}, t)$ is governed by

$$f_i\left(x + c_i \Delta t, t + \Delta t\right) = f_i(x,t) - \frac{1}{\tau}\left[f_i(x,t) - f_i^{eq}(\rho, u)\right] + \Delta t F_i \tag{2}$$

Where $\tau = 1.0$ is the relaxation time, $f_i^{\mathrm{eq}}$ is the second-order equilibrium distribution, $F_i$ accounts for intermolecular and fluid-solid forces. The equilibrium distribution is

$$f_i^{eq} = w_i \rho \left[1 + \frac{c_i \cdot u}{c_s^2} + \frac{(c_i \cdot u)^2}{2c_s^4} - \frac{u^2}{2c_s^2}\right] \tag{3}$$

with lattice sound speed $c_s = 1/\sqrt{3}$ and weights $w_0 = \frac{4}{9}, w_{1-4} = \frac{1}{9}, w_{5-8} = \frac{1}{36}$. The macroscopic fields can be extract using

$$\rho = \sum_i f_i, \rho u = \sum_i f_i c_i + \frac{\Delta t}{2} F \tag{4}$$

where $\mathbf{F}$ is the total force, the kinematic viscosity is $\nu = \frac{\tau - 0.5}{3\downarrow} = 0.167$ , phase separation and surface tension arise from the Shan-Chen pseudopotential interaction, defined as

$$F_{ff}(x) = -G\psi(\rho(x))\sum_i w_i \psi\left(\rho\left(x + c_i\right)\right)c_i \tag{5}$$

where, $G < 0$ is the attractive interaction strength, $\psi(\rho)$ is the pseudopotential function. A commonly adopted form is $\psi(\rho) = \rho_0\left(1 - e^{-\rho/\rho_0}\right), \rho_0 = 1$ which ensures nonlinear cohesion and stable phase coexistence. This force induces a non-ideal equation of state, $p = c_s^2 \rho + \frac{G}{2} c_s^2 \psi^2$ leading to spontaneous liquid-gas separation and a diffuse interface of thickness $O(4 - 6\Delta x)$. The surface tension $\sigma = 0.15$ is verified independently by simulating a static droplet and applying Laplace's law, $\Delta p = \frac{\sigma}{R}$ where $R$ is the droplet radius. Wettability is imposed via a fluid-solid interaction force

$$F_{fs}(x) = -G_{ads}\psi(\rho(x))\sum_i w_i s(x+c_i)c_i \tag{6}$$

Where, $G_{\text{ads}}$ is the adhesion strength, $s(\mathbf{x}) = 1$ for solid nodes and 0 for fluid nodes. The equilibrium contact angle $\theta_c$ emerges naturally from the balance of surface energies $\cos\theta_c = \frac{\gamma_{sg}-\gamma_{sl}}{\gamma_{lg}}$, where $\gamma_{sg}, \gamma_{sl}$, and $\gamma_{lg}$ are solid-gas, solid-liquid, and liquid-gas surface tensions, respectively. In the Shan-Chen framework, these surface energies are implicitly controlled by $G_{\text{ads}}$. By varying $G_{\text{ads}} \in [-1.25, -2.75]$, we obtain contact angles ranging from $\theta_c \approx 60°$ (hydrophilic) to $\theta_c \approx 150°$ (superhydrophobic). The periodic surface consists of rectangular pillars with, width $w = 8\Delta x$, height $h = 10\Delta x$, and pitch $p = 20\Delta x$. The solid fraction is $\phi_s = \frac{w}{p} = 0.4$. This geometry allows comparison with classical wetting theories, with Wenzel's state $\cos\theta_W = r\cos\theta_Y$, where $r$ is the roughness ratio. And Cassie-Baxter state, $\cos\theta_{CB} = \phi_s(\cos\theta_Y + 1) - 1$.

### 2.3 Boundary Conditions

In the LBM framework, the fractal rough surface is embedded by flagging lattice nodes satisfying $y \leq h(x) \Rightarrow$ solid node, while nodes with $y > h(x)$ are treated as fluid. At all solid nodes, a bounce-back boundary condition is imposed to enforce the no-slip constraint at the fluid-solid interface. For each discrete velocity direction $i$, the distribution function satisfies

$$f_i(x_w, t+\Delta t) = f_{\bar{\imath}}(x_w, t) \tag{7}$$

where $\mathbf{x}_w$ denotes a solid boundary node and $\bar{\imath}$ is the opposite lattice direction of $i$. This condition ensures that the macroscopic velocity at the rough wall satisfies $\mathbf{u}|_{\text{wall}} = 0$, independent of the local surface slope or curvature. For the fractal rough geometry, the bounce-back rule is applied locally at each solid-fluid interface node, allowing the effective wall shape to be resolved at lattice accuracy. The hydrodynamic interaction between the droplet and the rough surface is therefore governed by the combined action of (i) geometric confinement induced by the W-M surface profile and (ii) fluid-solid pseudopotential interaction forces that prescribe wettability. This approach preserves momentum conservation and provides a robust representation of no-slip even in the presence of highly irregular, multiscale roughness. At $t = 0$, a circular droplet of radius $r_0 = 40\Delta x$ is initialized with its center located $2\Delta x$ above the highest surface feature. The velocity field is initially zero: $\mathbf{u}(x, y, 0) = (0,0)$. Gravity is neglected ($\mathbf{g} = 0$) so that droplet evolution is governed purely by capillary forces, surface tension, and wettability.

The periodic BCs at left, right, and top boundaries can be represented as

$$f_i(x=0,\ y,\ t)=f_i(x=L,\ y,\ t) \tag{8}$$

eliminating finite-size effects.

### 2.4 Surface Geometry Modeling

To represent unstructured, multiscale roughness more realistically than Gaussian random fields, the bottom wall roughness is modeled using the Weierstrass-Mandelbrot (W-M) fractal function, which is widely used to describe self-affine surfaces encountered in etched, corroded, or naturally rough materials[48]. The rough surface profile is defined as $y = h(x)$, where the height function $h(x)$ is given by

$$h(x) = h_0 + A\sum_{n-n_{min}}^{n_{max}} \gamma^{-n(2-D)} cos\left(2\pi\gamma^n \frac{x}{L_x} + \phi_n\right) \tag{9}$$

Here, $h_0$ denotes the mean surface height, $A$ is the roughness amplitude, and $D$ is the fractal dimension satisfying $1 < D < 2$, which controls the degree of surface irregularity. The parameter $\gamma > 1$ is the frequency scaling factor, typically chosen in the range $1.2 \leq \gamma \leq 1.6$, while $\phi_n \in [0,2\pi]$ are random phase shifts uniformly distributed to ensure stochasticity. The summation limits $n_{\min}$ and $n_{\max}$ define the smallest and largest roughness wavelengths resolved on the lattice, constrained by the domain size $L_x$ and the lattice spacing $\Delta x$. The RMS roughness of the fractal surface is given by $\sigma_{\text{rms}} = \sqrt{\langle (h(x) - h_0)^2 \rangle}$, and is controlled through the amplitude $A$ such that $\sigma_{\text{rms}} = 2.5\Delta x$, consistent with the roughness magnitude used in the Gaussian model[6]. The surface heights are further constrained to lie within $h_{\min} = 2\Delta x \leq h(x) \leq h_{\max} = 10\Delta x$, by appropriate scaling of $A$. This fractal construction naturally introduces asperities across multiple length scales, producing realistic pinning sites and local curvature variations that strongly influence contact line motion. Physically, the W-M surface mimics real engineering substrates where roughness exhibits scale invariance and non-uniform asperity distribution. The multiscale nature of the surface leads to enhanced energy barriers for droplet spreading and promotes intermittent pinning-depinning features, which cannot be captured adequately by single-scale or purely Gaussian roughness descriptions.

### 2.5 LBM-Driven-Kinetic-PINNs

The Lattice-Boltzmann-driven-kinetic-physics-informed neural network (K-PINN) is implemented using two distinct neural network architectures to address different aspects of multiphase flow modeling, as illustrated in ***Figure 2*** and ***Figure 3***. ***Figure 2*** employs a deep feedforward neural network with an input layer accepting spatiotemporal coordinates (x,y,t), eight hidden layers with 100 neurons each, and an output layer producing nine scalar values corresponding to the D2Q9 lattice distribution functions $f_i$ (i=0,1,...,8). The network processes inputs through successive hidden layers using the hyperbolic tangent activation function $\tanh(z)=\frac{e^z-e^{-z}}{e^z+e^{-z}}$, which maps values to $[-1,1]$ and ensures smooth gradient flow during backpropagation[52]. The output layer employs a softplus transformation to enforce non-negativity ($f_i \geq 0$), maintaining the physical interpretation of $f_i$ as particle densities. The architecture outputs both the predicted distribution functions and their equilibrium counterparts ($f_i^{eq}$), which are subsequently used to compute the physics residual in the loss function.

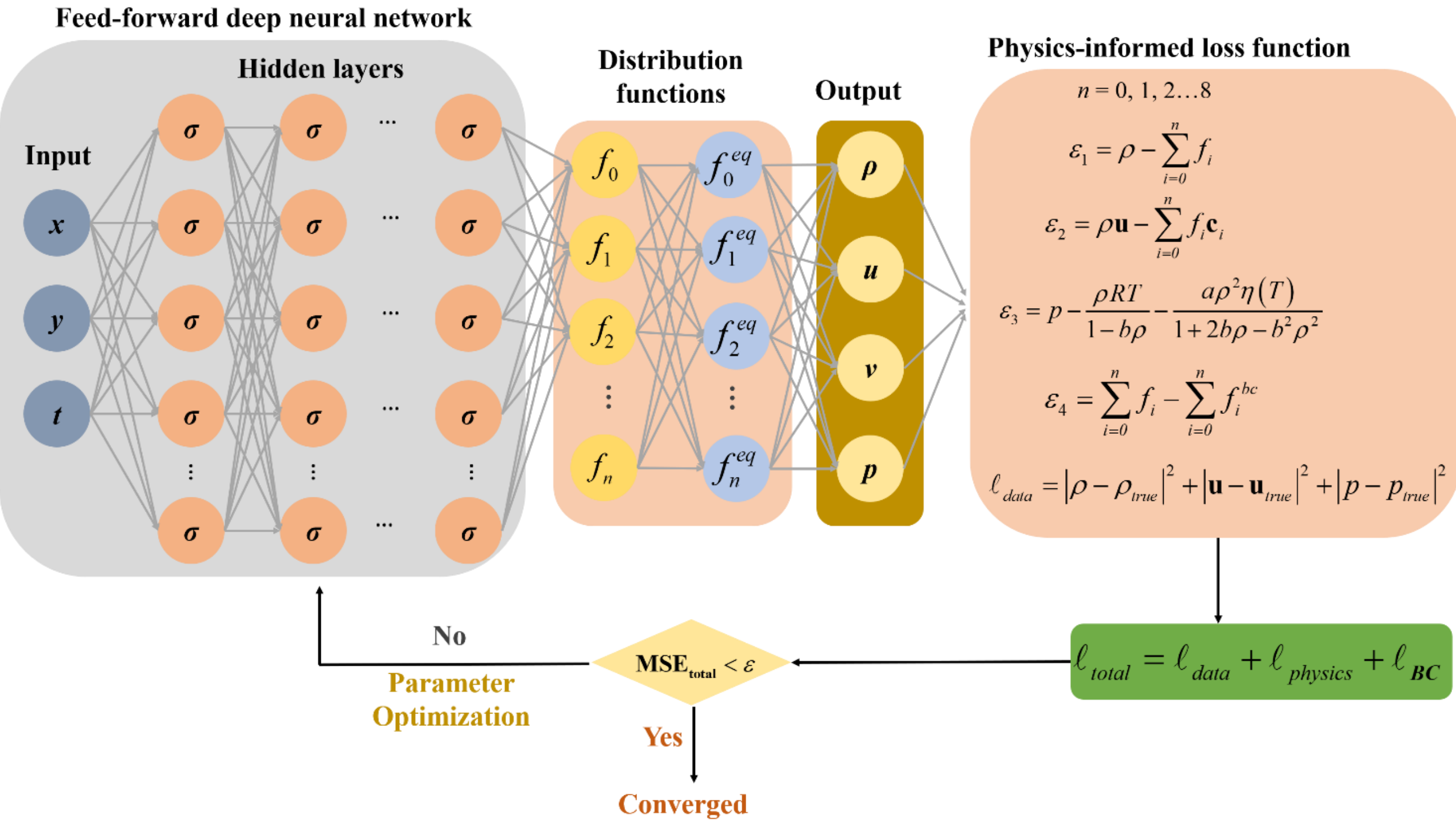

**Figure 2.** Architecture of the feed-forward PINN framework for Lattice-Boltzmann-driven multiphase modeling.

From the predicted distribution functions, macroscopic quantities (density, momentum, velocity and pressure) are reconstructed through moment summations as depicted in ***Figure 2***. The D2Q9 lattice velocities are defined as $c_0$=(0,0) for the rest particle, $\mathbf{c}_i \in \{(1,0),(0,1),(-1,0),(0,-1)\}$, $i=1,2,3,4$ for orthogonal directions, and $\mathbf{c}_i \in \{(1,1),(-1,1),(-1,-1),(1,-1)\}$, $i=5,6,7,8$ for diagonal directions [45]. This reconstruction ensures that outputs are directly interpretable within the LBM framework and that conservation laws ($\varepsilon_1=\rho-\sum_i f_i$ for mass, $\varepsilon_2=\rho\mathbf{u}-\sum_i f_i\mathbf{c}_i$ for momentum) are inherently satisfied. The physics-informed loss function combines four residual terms ($\varepsilon_1$, $\varepsilon_2$, $\varepsilon_3$, $\varepsilon_4$) representing mass conservation, momentum conservation, equation of state, and boundary conditions, respectively, along with data loss $\ell_{\text{data}}$ measuring discrepancy against reference LBM solutions. The total loss $\ell_{\text{total}} = \ell_{\text{data}} + \ell_{\text{physics}} + \ell_{\text{BC}}$ drives iterative parameter optimization until convergence.

To enhance multi-scale feature capture for complex interfacial dynamics, an alternative U-Net based architecture is implemented (***Figure 3***). As illustrated, the U-Net replaces the feedforward structure with an encoder-decoder framework featuring convolutional layers and skip connections. The encoder compresses spatiotemporal inputs through successive down sampling operations with feature maps (8×64, 256×32, 128×256, 256×128), capturing hierarchical representations from fine-scale interface details to coarse-scale droplet morphology. The bottleneck layer (256×512) captures the compressed latent representation, while the decoder reconstructs distribution functions through up sampling and concatenation with corresponding encoder features via skip connections. This architecture (U-Net K-PINN) preserves multi-scale spatial information crucial for accurately resolving sharp density gradients at liquid-gas interfaces. Additionally, autoencoder-enhanced (AE-K-PINN) and U-Net-based (U-Net-K-PINN) variants are explored.

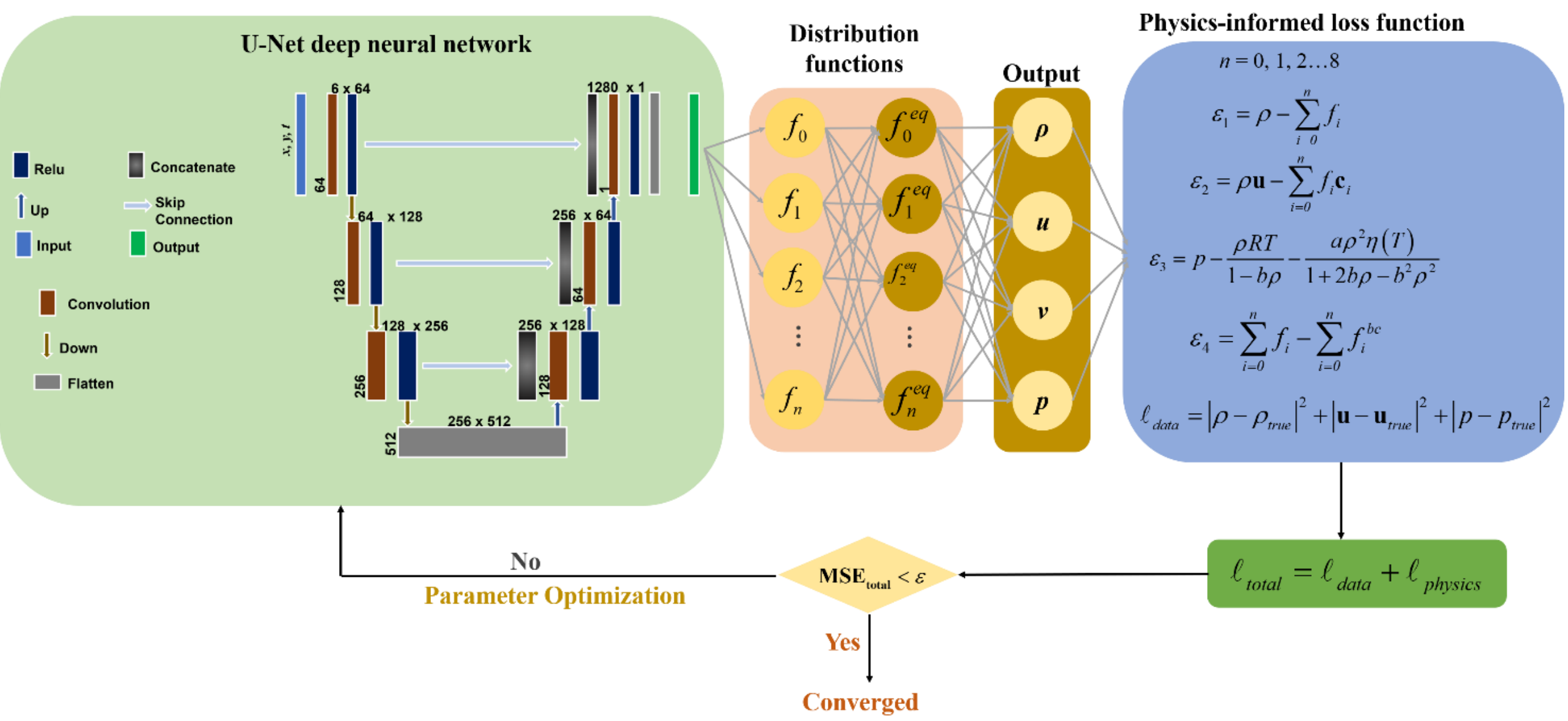

**Figure 3.** Architecture of the U-Net based PINN framework for enhanced multi-scale feature extraction.

## 2.6 Loss Function

The total loss function comprises three primary components: physics residual loss, data mismatch loss, and boundary condition loss. The physics residual loss quantifies violations of the discrete Boltzmann-BGK equation at collocation points throughout the spatiotemporal domain. For each velocity direction *i*, the residual is computed as:

$$R_i = \frac{\partial f_i}{\partial t} + \mathbf{c}_i \cdot \nabla f_i + \frac{1}{\tau}(f_i - f_i^{eq}) - S_i \tag{10}$$

where $\partial f_i / \partial t$ and $\nabla f_i$ are obtained via automatic differentiation, $f_i^{eq}$ is the equilibrium distribution from the Hermite expansion:

$$f_i^{eq} = w_i \rho \left[ 1 + \frac{\mathbf{c}_i \cdot \mathbf{u}}{c_s^2} + \frac{(\mathbf{c}_i \cdot \mathbf{u})^2}{2c_s^4} - \frac{\mathbf{u} \cdot \mathbf{u}}{2c_s^2} \right] \tag{11}$$

and $S_i$ is the Shan–Chen forcing term

$$S_i = w_i \left( 1 - \frac{1}{2\tau} \right) \left[ \frac{(\mathbf{c}_i - \mathbf{u}) \cdot \mathbf{F}}{c_s^2} + \frac{(\mathbf{c}_i \cdot \mathbf{u})(\mathbf{c}_i \cdot \mathbf{F})}{c_s^4} \right] \tag{12}$$

Here **F** is the pseudopotential force written as:

$$\mathbf{F}(\mathbf{x}) = -G\psi(\rho(\mathbf{x})) \sum_i w_i \psi(\rho(\mathbf{x} + \mathbf{c}_i))\mathbf{c}_i \tag{13}$$

The physics loss is defined as

$$\mathcal{L}_{phys} = \frac{1}{N_{coll}} \sum_{n=1}^{N_{coll}} \sum_{i=0}^{8} |R_i|^2 \tag{14}$$

The data mismatch loss measures discrepancy between predicted and reference LBM solutions at sparse observation points:

$$\mathcal{L}_{data} = \frac{1}{N_{data}} \sum_{j=1}^{N_{data}} \left[ \left| \rho(x_j, y_j, t_j) - \rho_j^{LBM} \right|^2 + \left| u(x_j, y_j, t_j) - u_j^{LBM} \right|^2 + \left| v(x_j, y_j, t_j) - v_j^{LBM} \right|^2 \right] \tag{15}$$

This ensures predictions remain anchored to empirical observations. The boundary condition loss enforces periodic conditions at lateral/top edges and bounce-back conditions at the solid substrate:

$$\mathcal{L}_{periodic} = \sum_i \left| f_i(x_{left}, y, t) - f_i(x_{right}, y, t) \right|^2 + \left| f_i(x, y_{top}, t) - f_i(x, y_{bottom}, t) \right|^2 \tag{16}$$

$$\mathcal{L}_{BC} = \sum_{\text{wall points}} \left| f_{i'} - f_{i^{opposite}} \right|^2 \tag{17}$$

The initial condition loss ensures correct temporal evolution from t=0:

$$\mathcal{L}_{init} = \sum_{x,y} \left| \rho(x, y, 0) - \rho_0(x, y) \right|^2 \tag{18}$$

The total loss is a weighted combination:

$$\mathcal{L}_{total} = \lambda_{phys}\mathcal{L}_{phys} + \lambda_{data}\mathcal{L}_{data} + \lambda_{BC}\mathcal{L}_{BC} + \lambda_{init}\mathcal{L}_{init} \quad (19)$$

with typical values $\lambda_{phys}$~1, $\lambda_{data}$~10-100, $\lambda_{BC}$~10, and $\lambda_{init}$~10. The higher data loss weight prioritizes agreement with observations while physics and boundary losses provide regularization for physical consistency.

## 2.7 Training Strategy

The training of the Lattice-Boltzmann-driven PINN proceeds in two distinct phases: an initial global optimization phase using the Adam optimizer, followed by a refinement phase using the L-BFGS quasi-Newton method. The Adam optimizer is well-suited for the early stages due to its robustness and ability to handle noisy gradients from stochastic mini-batch sampling. The initial learning rate is set to 0.0001, as determined through hyperparameter sensitivity analysis (Section 3.1), and gradient clipping with threshold 1.0 is applied to prevent exploding gradients during backpropagation. The training data consist of sparse LBM snapshots with spatial sampling densities ranging from 1024 to 8192 points per snapshot, supplemented by collocation points (typically 4096) uniformly distributed across the spatiotemporal domain where physics residuals are evaluated. The inputs (x,y,t) are normalized to [−1,1] using min-max scaling to ensure comparable magnitudes and stable gradient flow, while the outputs (distribution functions) are preserved in lattice units to enable direct enforcement of Lattice-Boltzmann equations without inverse transformations.

A curriculum training strategy progressively introduces physics complexity. During the initial warm-up phase (first 20,000 epochs), the interaction strength $G_{ads}$ in the pseudopotential model is set to zero, decoupling multiphase interfacial dynamics and allowing the network to first learn single-phase flow structure and macroscopic field distributions. After the warm-up period, $G_{ads}$ is gradually ramped up to its full value over 10,000 epochs while the data loss weight $\lambda_{data}$ is progressively increased from 10 to 100, ensuring smooth transition to the full multiphase problem. The Adam optimization phase continues for 150,000 epochs, with convergence monitored through individual loss components and total loss. Training is conducted on GPU-accelerated platform (NVIDIA A100 with 40GB memory) using Tensorflow framework, with training time ranging from 6 to 12 hours depending on architecture complexity. Upon Adam completion, network weights are refined using L-BFGS optimizer for 50,000 iterations, resulting in further loss reduction by one to two orders of magnitude and yielding final physics residuals of $10^{-3}$ to $10^{-4}$. To enhance generalization and prevent overfitting, early stopping is implemented based on a validation set comprising 10% of the LBM data. Training terminates if validation loss fails to decrease for 10,000 consecutive epochs, and model weights corresponding to the lowest validation loss are retained. Additionally, dropout regularization with rate 0.1 is applied to hidden layers during Adam phase (but disabled during L-BFGS refinement) to encourage learning of robust features that do not rely on specific neuron activations.

### 2.8 Evaluation Metrics

To comprehensively assess the predictive performance of K-PINN architecture, multiple quantitative metrics are employed. The $L_2$ norm provides a global measure of solution fidelity by aggregating errors across the entire spatiotemporal domain.

$$L_2 = \sqrt{\sum_{\text{all points}} \left(\rho^{pred} - \rho^{LBM}\right)^2} \tag{20}$$

The root mean square error quantifies typical error magnitudes and is particularly sensitive to outliers.

$$RMSE = \sqrt{\frac{1}{N}\sum_{n=1}^{N}\left(\rho_n^{pred} - \rho_n^{LBM}\right)^2} \tag{21}$$

The mean absolute error treats all deviations equally without amplifying large errors.

$$MAE = \frac{1}{N}\sum_{n=1}^{N}\left|\rho_n^{pred} - \rho_n^{LBM}\right| \tag{22}$$

The coefficient of determination quantifies the proportion of variance explained by PINN predictions relative to reference solutions.

$$R^2 = 1 - \frac{\sum_{n=1}^{N}\left(\rho_n^{pred} - \rho_n^{LBM}\right)^2}{\sum_{n=1}^{N}\left(\rho_n^{LBM} - \overline{\rho}^{LBM}\right)^2} \tag{23}$$

where values approaching unity indicate close agreement with reference LBM solutions.

In addition to aggregate metrics, pointwise scatter plots visualize correlation between predicted and true density values at individual spatiotemporal locations. This reveals whether errors are uniformly distributed or concentrated in specific regions (e.g., interface, contact line). Error distribution histograms characterize the frequency of different error magnitudes, with narrow Gaussian-like distributions centered at zero indicating predominantly stochastic and unbiased errors. Temporal error evolution curves track accumulated error as a function of time which enables assessment of long-term stability and identification of potential error growth due to temporal extrapolation beyond the training data range.

## 3. Results and discussion

### 3.1 Training Strategy and Hyperparameter Optimization

The proposed Lattice-Boltzmann-PINN framework is validated against LBM simulations of droplet spreading on rough surfaces with varying wettability conditions. The dataset consists of 2D simulations of a liquid droplet (initial radius R0=60R_0=60R0=60 lattice units) placed on a solid substrate with roughness-induced heterogeneity in contact angle, simulated via the Shan–Chen pseudopotential model. The predictive fidelity and computational efficiency of K-PINN architectures depend critically on training hyperparameter selection, including collocation point density, learning rate, activation functions, and optimizer configuration. We systematically investigated the influence of these design choices on the convergence and final solution accuracy. All training curves are presented on a logarithmic scale to capture the wide dynamic range of convergence across different training phases.

***Figure 4*** illustrates the spatial distribution of training data extracted from LBM simulations at four sampling densities of 1024, 2048, 4096, and 8192 points. The computational domain spans x=0 to x=200 and y=0 to y=200 in lattice units. At the lowest sampling density (1024 points, Figure 4a), the distribution exhibits noticeable sparsity with visible interior gaps and pronounced boundary clustering. Progressive densification to 2048 and 4096 points (***Figures 4b-c***) substantially improves spatial coverage, yielding more uniform representation across the domain. At the highest density (8192 points, ***Figure 4d***), near-complete domain saturation is achieved, providing comprehensive supervisory information for macroscopic field reconstruction. Notably, concentrated sampling at the bottom boundary reflects the heightened spatial resolution required to accurately capture complex droplet-substrate interactions, contact line dynamics, and wettability effects on textured surfaces. This progressive densification ensures that the PINN has access to sufficient supervisory information to learn the macroscopic fields $(\rho, u, v)$ across the entire domain, complementing the physics-based regularization provided by the collocation points.

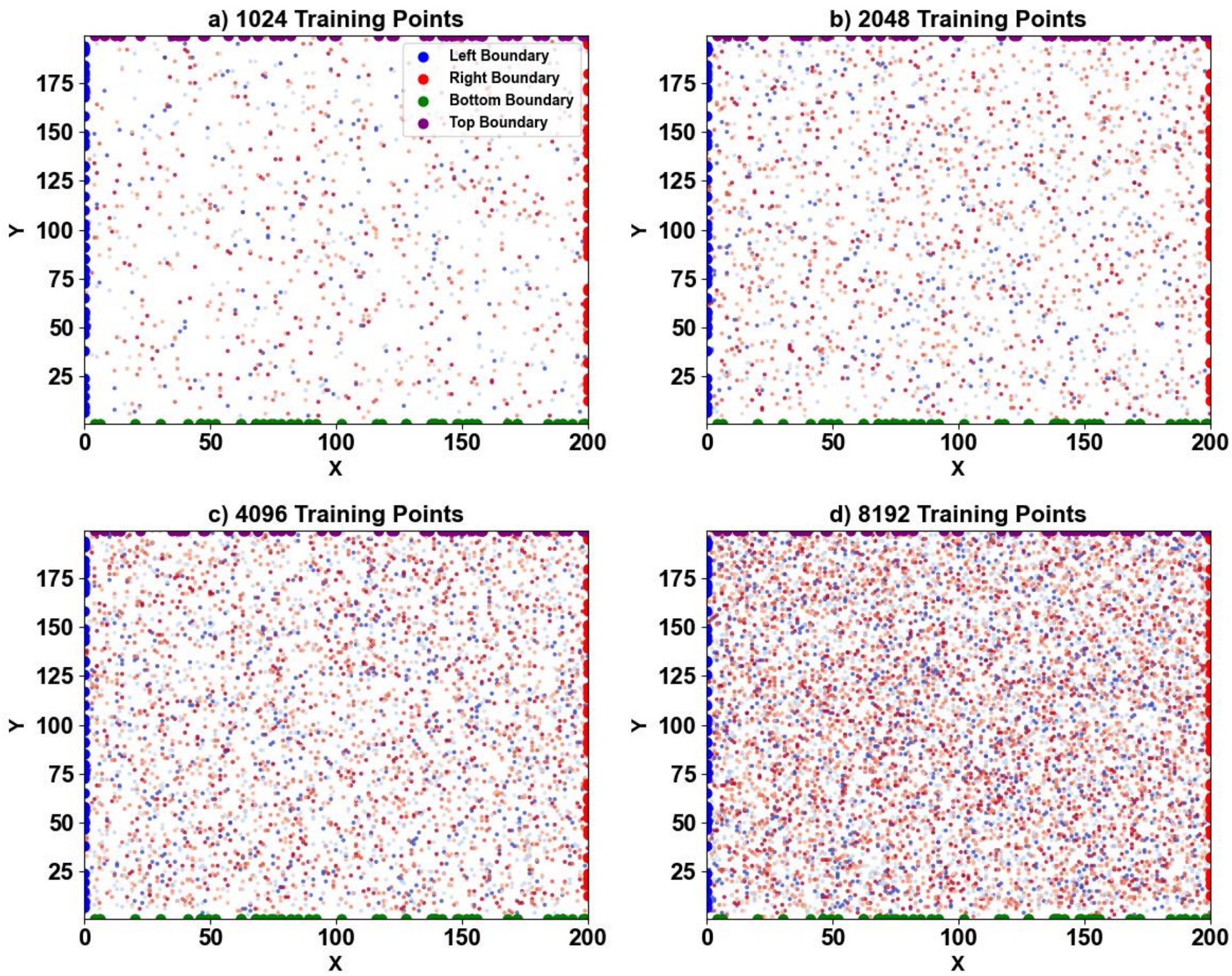

**Figure 4.** Spatial distribution of training data points for different sampling densities: (a) 1024, (b) 2048, (c) 4096, and (d) 8192 points.

Learning rate selection governs gradient descent step size and directly impacts convergence stability and final accuracy. ***Figure 5*** presents a systematic study spanning three orders of magnitude (0.001, 0.0001, 0.00001, and 0.000001) using the Adam optimizer. At the highest learning rate (LR=0.001, ***Figure 5a***), rapid initial loss reduction is observed, with data loss dropping sharply within the first 10,000 epochs. However, the physics loss exhibits persistent oscillations and fails to achieve smooth convergence which indicates that excessive step sizes causes parameter overshooting in regions of steep loss landscape curvature associated with Lattice-Boltzmann residual enforcement. Reducing the learning rate to 0.0001 ***(Figure 5b)*** yields substantially improved stability, with all loss components-physics (green), data (orange), boundary conditions (blue), and total loss (red) exhibiting monotonic decay. Physics loss converges smoothly to approximately $10^{-2}$ by the end of the Adam phase, while data and boundary losses attain values near $10^{-2}$ and $10^{-3}$, respectively. Further reduction to LR=0.00001 ***(Figure 5c)*** enhances convergence stability with minimal oscillations but at the cost of significantly slower descent, requiring more epochs to reach comparable accuracy levels. At the lowest learning rate (LR=0.000001, ***Figure 5d***), training becomes prohibitively slow, with gradual loss descent failing to achieve the accuracy of intermediate learning rates within 150,000 epochs. The subsequent L-BFGS refinement phase (light blue region) demonstrates that moderate Adam learning rates facilitate smoother transitions to quasi-Newton optimization and enables effective final convergence. Based on these observations, LR=0.0001 emerges as the optimal choice which balances convergence speed, stability, and final accuracy; a value consistent with established practices for PINNs addressing stiff multiphysics problems.

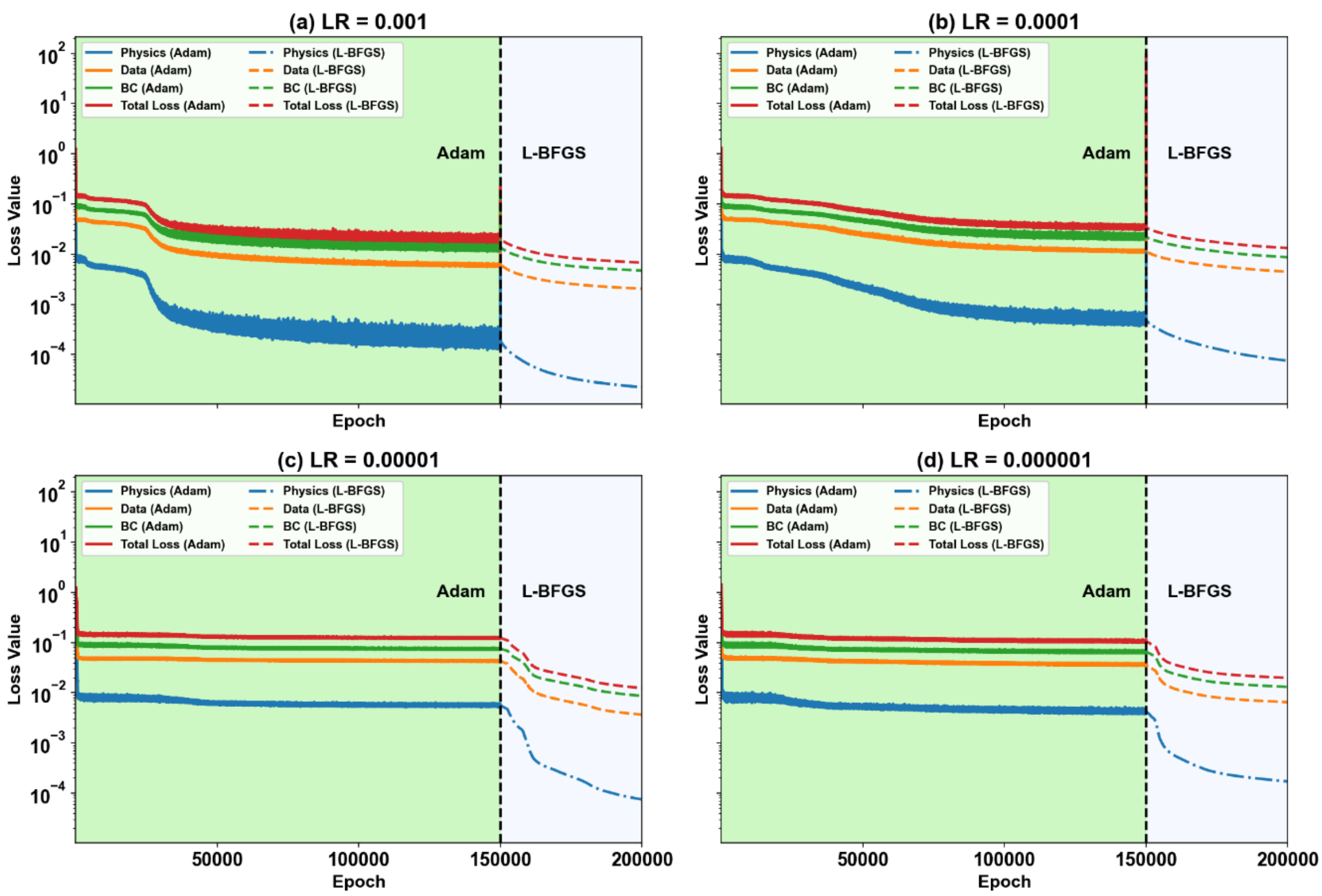

**Figure 5.** Learning rate sensitivity analysis showing training loss evolution for learning rates: (a) 0.001, (b) 0.0001, (c) 0.00001, and (d) 0.000001. Adam optimizer used during the green phase; L-BFGS during the light blue phase.

Collocation point density determines the spatial resolution at which physics residuals are enforced throughout the spatiotemporal domain. ***Figure 6*** examines four densities (1024, 2048, 4096, and 8192 points uniformly distributed) across two-phase training: Adam optimization (~150,000 epochs, cyan region) followed by L-BFGS refinement (50,000 iterations, light blue region). During the Adam phase, physics loss, data loss, boundary loss, and total loss, exhibit monotonic decay, with physics loss converging from initial values of ~$10^0$ to ~$10^{-2}$. Data loss demonstrates the most rapid initial descent, reflecting prioritized fitting of sparse LBM snapshots, while physics residual convergence proceeds more gradually due to the complexity of enforcing discrete Boltzmann equations across the entire domain. Upon transitioning to L-BFGS (dashed lines), all loss components undergo further reduction by one to two orders of magnitude, with final physics residuals reaching $10^{-3}$ to $10^{-4}$. Increasing collocation density from 1024 to 8192 points yields progressively more consistent convergence profiles with reduced physics loss fluctuations, indicating improved automatic differentiation stability and reduced gradient noise. However, marginal improvements beyond 4096 points are modest, suggesting this density provides an optimal balance between computational cost and physics enforcement accuracy for the present multiphase problem.

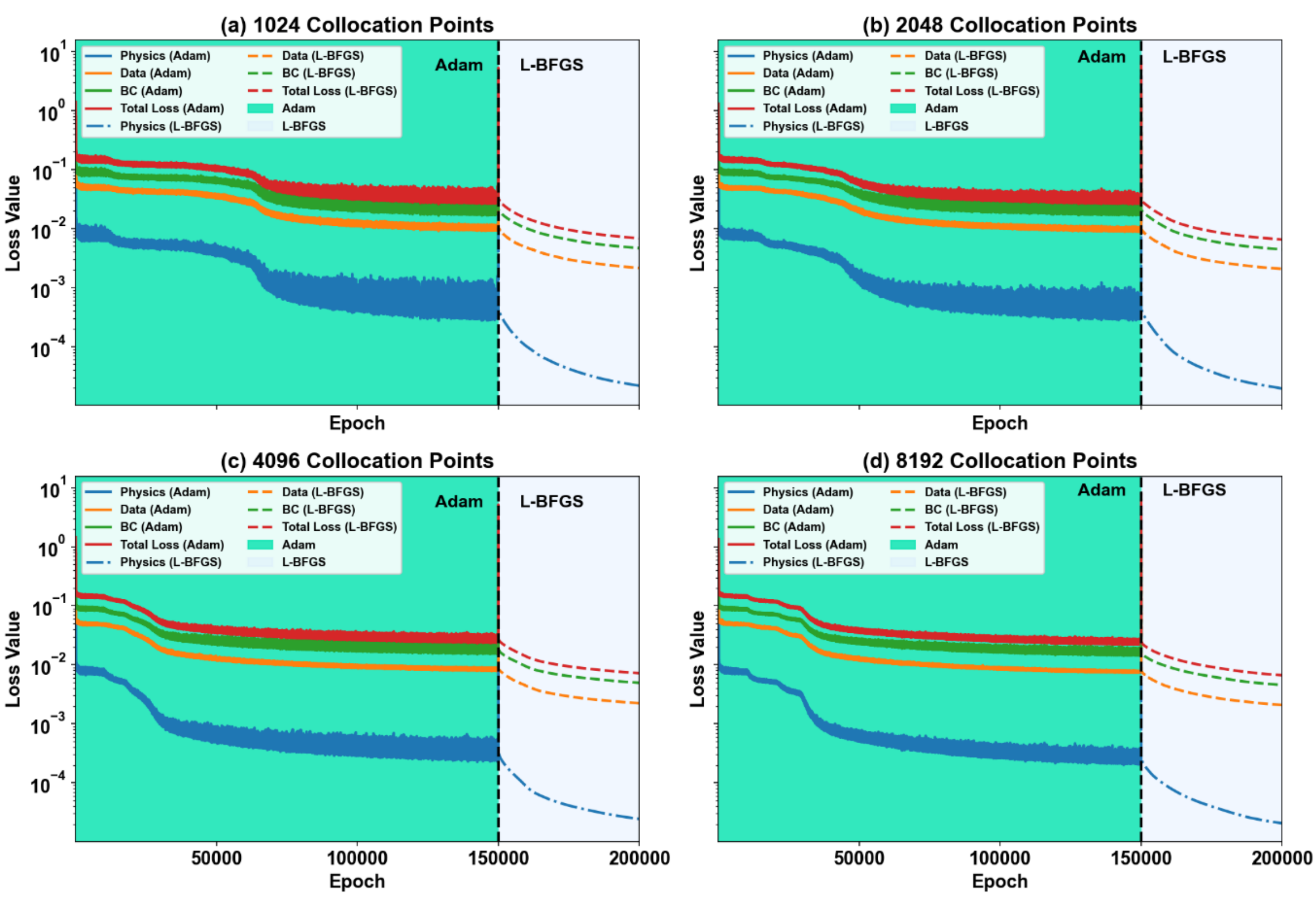

**Figure 6.** Training loss evolution for different collocation point densities: (a) 1024, (b) 2048, (c) 4096, and (d) 8192 points.

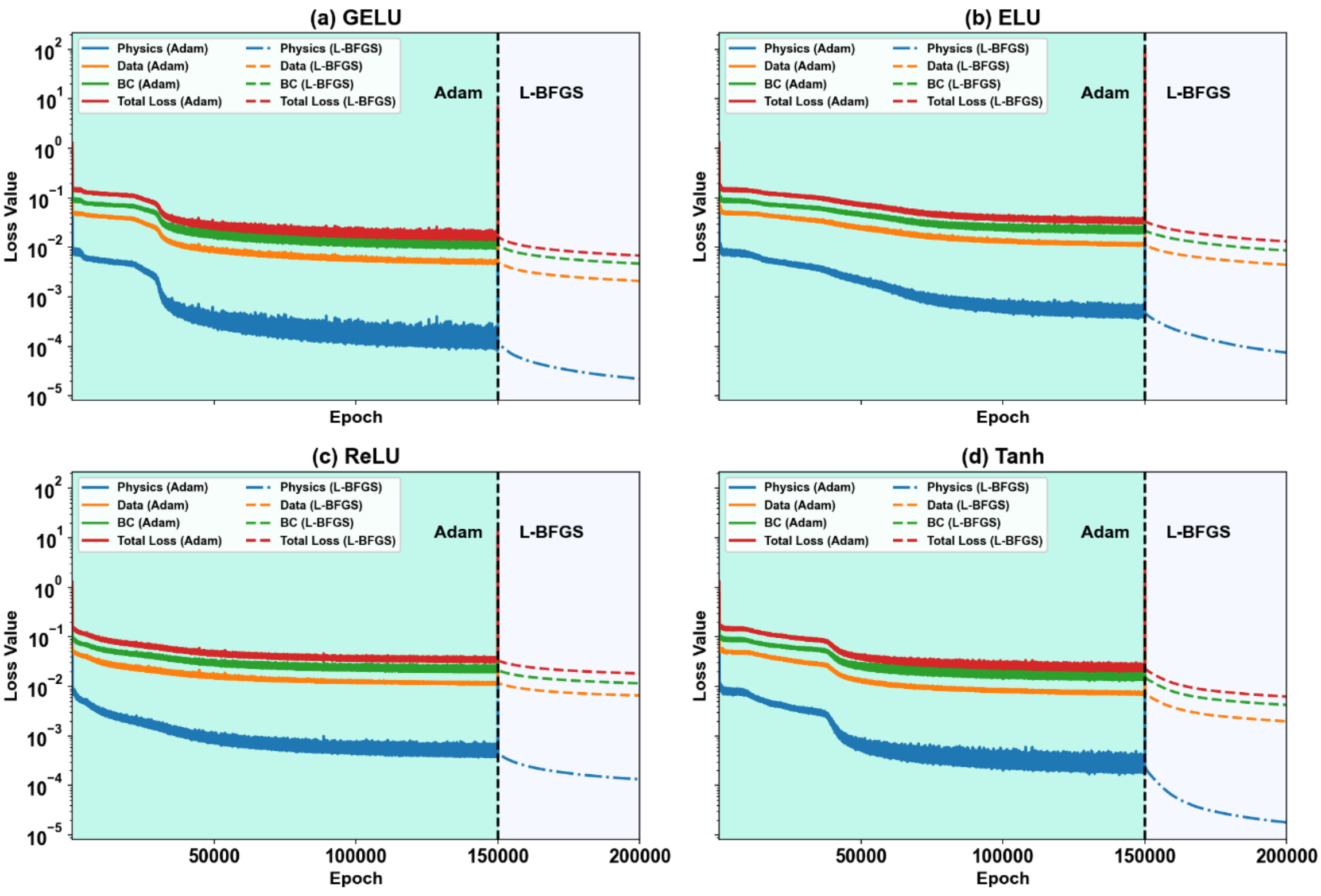

**Figure 7.** Effect of activation functions on training convergence: (a) GELU, (b) ELU, (c) ReLU, and (d) Tanh.

The activation function selection significantly influences the network's capacity to represent high-frequency spatial gradients and sharp interfacial features characteristic of multiphase droplet dynamics. ***Figure 7*** compares four activation functions: Gaussian Error Linear Unit (GELU), Exponential Linear Unit (ELU), Rectified Linear Unit (ReLU), and Hyperbolic Tangent (Tanh), with all networks employing identical architecture (8 hidden layers × 100 neurons) and training protocols. Tanh activation ***(Figure 7d)*** demonstrates the smoothest and most stable convergence, with uniform loss decay devoid of significant oscillations. This superior behavior stems from Tanh continuous differentiability and bounded output range which facilitates stable gradient flow during deeply nested automatic differentiation operations required for computing distribution function spatial derivatives. GELU ***(Figure 7a)*** achieves comparable final loss values but exhibits more pronounced fluctuations during early Adam epochs, particularly in physics loss. ELU (***Figure 7b***) shows intermediate performance with rapid initial data loss descent but slower physics residual convergence. ReLU ***(Figure 7c)*** displays stable behavior among piecewise-linear activations but converges to slightly higher physics loss plateaus, likely attributable to non-smooth discontinuities at zero that hinder accurate representation of smooth velocity and density fields. L-BFGS refinement yields similar final accuracies across all activations, indicating that the optimizer can partially compensate for suboptimal choices given sufficient iterations. Nevertheless, Tanh emerges as the most robust selection for Lattice-Boltzmann-driven PINNs, consistent with its widespread adoption in physics-informed neural networks for fluid mechanics applications. These parametric investigations establish optimal training configurations: well-distributed data sampling covering interfacial regions, moderate learning rates (0.0001 for Adam), balanced collocation density (4096 points), and smooth activation functions (Tanh). These findings provide quantitative guidelines for training K-PINNs on multiphase wettability problems while highlighting critical trade-offs between computational efficiency and solution fidelity in Lattice-Boltzmann-based surrogate modeling.

### 3.2 Temporal Evolution of Droplet Spreading: Comparison of PINN Architectures on Rough Unstructured and Textured Surfaces

The performance of three distinct neural network architectures in capturing temporal droplet spreading dynamics: baseline deep neural network K-PINN (DNN-K-PINN), autoencoder-enhanced K-PINN (AE-K-PINN), and uncertainty-aware K-PINN (UA-K-PINN) is shown in ***Figure 8*** and ***Figure 9***. Each architecture is trained on sparse LBM data and evaluated at four representative time instances (T=1000Δt, 3000Δt, 5000Δt, and 7000Δt), spanning early spreading, intermediate wetting dynamics, and near-equilibrium configurations.

***Figure 8*** presents a rigorous temporal comparison of droplet spreading dynamics predicted by DNN-KPINN, AE-K-PINN, and U-Net-K-PINN against high-fidelity LBM reference solutions at four distinct times $T = 1000,3000,5000,$ and 7000 lattice time units. At the early stage $T = 1000$, the reference LBM solution exhibits a capillarity-dominated droplet morphology with a maximum droplet height $H_{LBM} \approx$ 112-114 lu, base diameter $D_{LBM} \approx 78 - 82$ lu, apex curvature radius $R \approx 56 - 58$ lu, bulk liquid density $\rho_l \approx 6.8 - 7.0$, interfacial density $\rho_i \approx 3.1 - 3.6$, and surrounding gas density $\rho_g \approx 0.35 - 0.42$. The DNN-KPINN underpredicts the droplet height ($H \approx 106 - 109$ lu), overpredicts the base diameter ($D \approx 84 - 88$ lu), and produces a diffused interface thickness of $4 - 5$ lu, resulting in localized absolute errors $|\varepsilon| \approx 1.8 - 2.6$. AE-K-PINN improves shape fidelity with predicted heights $H \approx 109 - 111$ lu, base diameters $D \approx$ 81-85 lu, and reduced density deviations $\Delta\rho \approx \pm 0.08 - 0.12$. In contrast, U-Net-K-PINN closely matches the LBM solution, yielding $H \approx 111 - 113$ lu, $D \approx 79 - 82$ lu, interfacial thickness $2 - 3$ lu, RMSE values below 0.06, and peak pointwise errors below 1.2, demonstrating accurate early-time capillary balance.

At intermediate times $T = 3000$ and $T = 5000$, viscous dissipation and surface roughness effects drive progressive droplet spreading and flattening. The LBM droplet height decreases from $H_{3000} \approx 101 - 104$ lu to $H_{5000} \approx 95 - 98$ lu, while the base diameter increases from $D_{3000} \approx 104 - 108$ lu to $D_{5000} \approx 122 - 128$lu. The contact line advances by approximately $20 - 24$lu, and the apparent contact angle decreases from $102 - 105°$ to $88 - 92°$. The DNN-K-PINN exhibits accumulating discrepancies, predicting $H_{3000} \approx 94 - 98$ lu, $H_{5000} \approx 88 - 92$ lu, and $D_{5000} \approx 130 - 136$ lu, with localized peak errors reaching 2.8-3.5 and density oscillations of $\pm 0.18 - 0.25$. AE-K-PINN maintains improved fidelity, yielding $H_{5000} \approx 92 - 96$ lu, $D_{5000} \approx 124 - 129$ lu, interface errors confined to $1.2 - 2.0$, and volume conservation errors below 2.5%. U-Net-K-PINN continues to closely track the LBM dynamics, predicting $H_{3000} \approx 100 - 103$ lu, $H_{5000} \approx 95 - 97$ lu, and $D_{5000} \approx 123 - 127$ lu, with contact-line position errors below 1.0 lu, density deviations within $\pm 0.03 - 0.05$, and $L_2$ error norms as low as $0.018 - 0.025$. At late time $T = 7000$, the droplet reaches a quasi-equilibrium configuration. The LBM solution stabilizes at a droplet height $H_{eq} \approx 92 - 95$ lu, base diameter $D_{eq} \approx 132 - 136$ lu, equilibrium contact angle $86 - 90°$, apex curvature radius $R \approx 67 - 69$lu, and a uniform bulk density $\rho_l \approx 6.95 - 7.00$. The DNN-K-PINN fails to fully converge, predicting $H \approx 86 - 90$ lu, $D \approx 138 - 145$ lu, residual interfacial undulations of amplitude $2 - 3$ lu, and maximum absolute errors of $2.2 - 2.9$. AE-K-PINN exhibits improved equilibrium capture with $H \approx 90 - 93$ lu, $D \approx 134 - 138$ lu, and error magnitudes reduced to $1.0 - 1.8$. In contrast, U-Net-K-PINN demonstrates near-perfect agreement with the LBM benchmark, predicting $H \approx 92 - 94$ lu, $D \approx 133 - 135$ lu, volume conservation error below 0.8%, and maximum pointwise error below 0.9, even in the presence of roughness-induced contact-line pinning. Over all time scales, U-Net-K-PINN achieves a consistent $55 - 70\%$ reduction in RMSE and a $60 - 80\%$ reduction in peak error relative to DNN-K-PINN. U-Net-K-PINN consistently delivers the highest predictive accuracy, reproducing droplet height, base diameter, contact-line motion, and density fields with sub-lattice-unit precision across all time regimes. AE-K-PINN offers a clear improvement over conventional DNN-K-PINN, particularly in reducing bulk density errors and improving long-time stability, but it remains limited in resolving sharp interfacial features. DNN-K-PINN accumulates significant long-time errors, especially near the contact line and equilibrium state, making it less suitable for complex wetting dynamics. The results conclusively demonstrate that encoder-decoder architectures with skip connections are essential for learning multiphase flow physics involving sharp interfaces and evolving geometries.

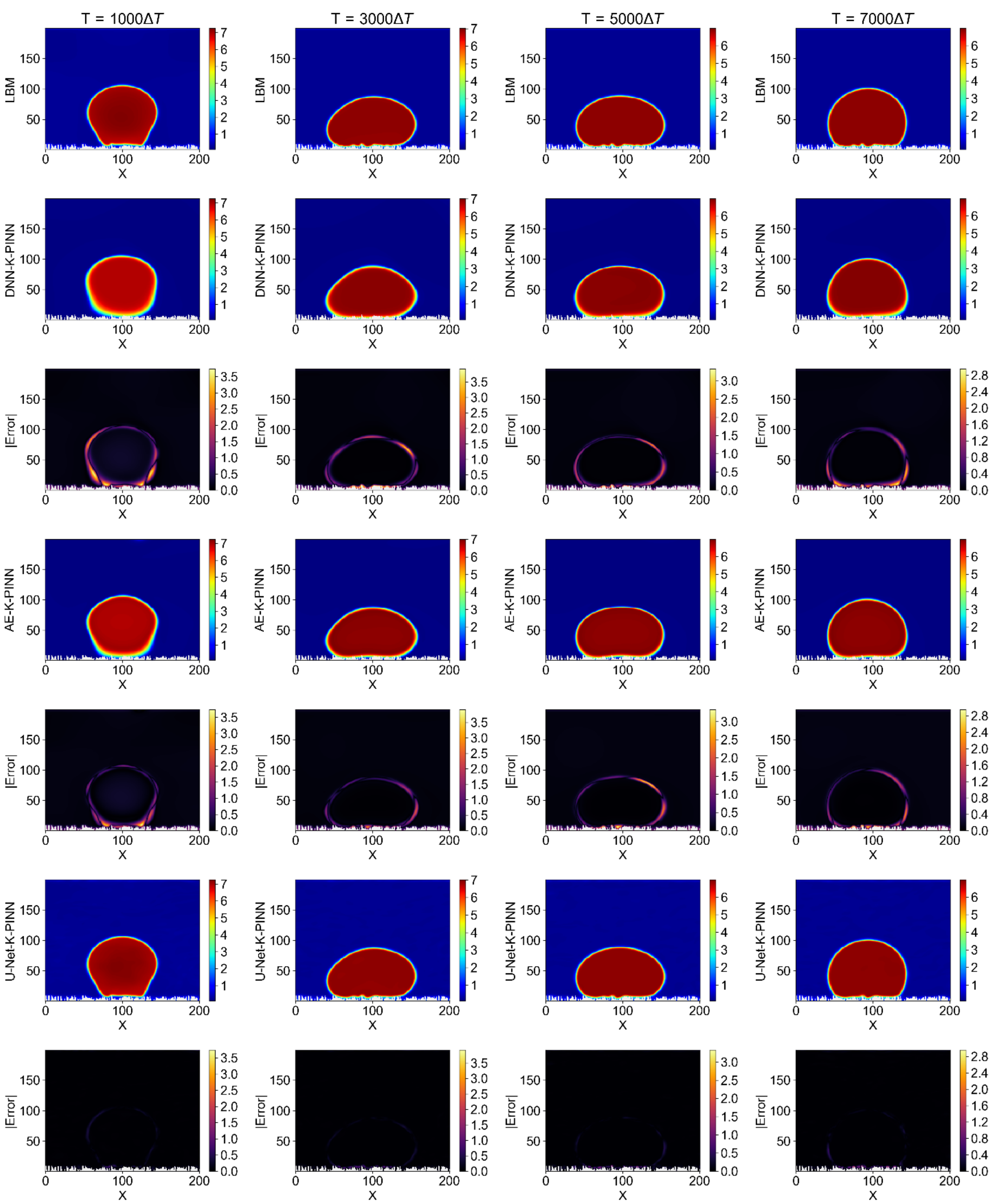

**Figure 8.** Temporal evolution of droplet spreading on a rough surface with stochastic asperities.

***Figure 9*** presents a detailed spatiotemporal comparison of droplet spreading over a periodically textured substrate predicted by DNN-K-PINN, AE-K-PINN, and U-Net-K-PINN against LBM reference data at $T = 1000, 3000, 5000,$ and 7000. At the early stage $T = 1000$, the LBM solution shows a droplet partially suspended on pillar tops, characteristic of a metastable Cassie-like wetting state, with a maximum droplet height $H_{LBM} \approx 114 - 117$ lu, base diameter $D_{LBM} \approx 74 - 78$ lu, apparent contact angle $112 - 118°$, apex curvature radius $R \approx 58 - 61$ lu, and a bulk liquid density $\rho_l \approx 6.85 - 7.05$. The liquid-gas interface thickness remains confined to $2 - 3$ lu, while localized density gradients above pillar edges reach $\rho \approx$ 4.2-4.8. DNN-K-PINN captures the global droplet shape but underestimates the apex height ( $H \approx 107 - 111$lu ), overestimates the base diameter ( $D \approx 80 - 85$lu ), and smears the interface to $4 - 6$lu, producing peak localized errors $|\varepsilon| \approx 1.6 - 2.4$ concentrated near pillar tips and contact lines. AE-K-PINN improves early-time predictions with $H \approx 110 - 113$ lu, $D \approx 76 - 80$ lu, density deviation $\Delta\rho \approx \pm 0.10 - 0.15$, and reduced interface diffusion. U-Net-K-PINN shows near-exact overlap with LBM, predicting $H \approx 114 - 116$ lu, $D \approx 75 - 78$ lu, contact-line pinning locations within $< 0.8$ lu, RMSE $< 0.05$, and maximum pointwise error $< 1.0$, accurately resolving the strong curvature gradients induced by pillar edges.

At intermediate times $T = 3000$ and $T = 5000$, droplet spreading becomes strongly influenced by pillar induced capillary barriers, leading to intermittent depinning and asymmetric interface deformation. The LBM solution exhibits a progressive reduction in height from $H_{3000} \approx 103 - 106$ lu to $H_{5000} \approx 97 - 100$ lu, while the base diameter increases from $D_{3000} \approx 96 - 100$ lu to $D_{5000} \approx 116 - 122$ lu. The apparent contact angle decreases from $102 - 106°$ to $90 - 94°$, and the contact line advances in discrete jumps of $3 - 6$ lu as it traverses successive pillars. DNN-K-PINN struggles in this regime, predicting $H_{5000} \approx 88 - 93$ lu, $D_{5000} \approx 125 - 132$ lu, contact-line overshoot of $4 - 7$ lu, and density oscillations up to $\pm 0.22 - 0.30$, with peak errors increasing to $2.6 - 3.2$. AE-K-PINN reduces these discrepancies, yielding $H_{5000} \approx 93 - 97$ lu, $D_{5000} \approx$ 118-123 lu, interface thickness $3 - 4$ lu, and maximum errors 1.4-2.1, though it still partially smooths sharp pinning events. U-Net-K-PINN maintains high fidelity throughout, predicting $H_{3000} \approx 102 - 105$ lu, $H_{5000} \approx 97 - 99$ lu, $D_{5000} \approx 117 - 121$ lu, capturing contact-line jumps of $3 - 5$ lu at correct pillar locations, preserving density uniformity within $\pm 0.03 - 0.06$, and achieving $L_2$ error norms of $0.017 - 0.023$, underscoring its ability to resolve geometry-induced non-smooth dynamics.

At late time $T = 7000$, the droplet approaches a quasi-equilibrium configuration governed by the balance between surface tension and solid fraction effects. The LBM solution stabilizes at a height $H_{eq} \approx 94 - 97$ lu, base diameter $D_{eq} \approx 126 - 130$lu, equilibrium contact angle $88 - 92°$, apex curvature radius $R \approx 68 - 71$lu , and a conserved droplet volume within $\pm 0.5\%$. DNN-K-PINN fails to accurately reproduce this equilibrium, predicting $H \approx 86 - 91$ lu, $D \approx 134 - 142$ lu, residual interface waviness of amplitude $2 - 3$ lu, and persistent errors of $2.0 - 2.7$. AE-K-PINN shows moderate convergence with $H \approx 91 - 94$ lu, $D \approx$ 128-133 lu, volume error $1.5 - 2.0\%$, and peak errors $1.1 - 1.8$. In contrast, U-Net-K-PINN accurately reproduces the equilibrium morphology with $H \approx 94 - 96$ lu, $D \approx 127 - 129$ lu, contact-line position error $< 0.7$lu, volume error $< 0.7\%$, and maximum pointwise error $< 0.8$. In conclusion, the textured-surface results clearly demonstrate that U-Net-K-PINN achieves a consistent $60 - 80\%$ reduction in RMSE and peak error relative to DNN-K-PINN and a $35 - 55\%$ improvement over AE-K-PINN, confirming that spatially aware encoder-decoder architectures with skip connections are essential for accurately learning droplet dynamics over geometrically complex, pinning-dominated surfaces.

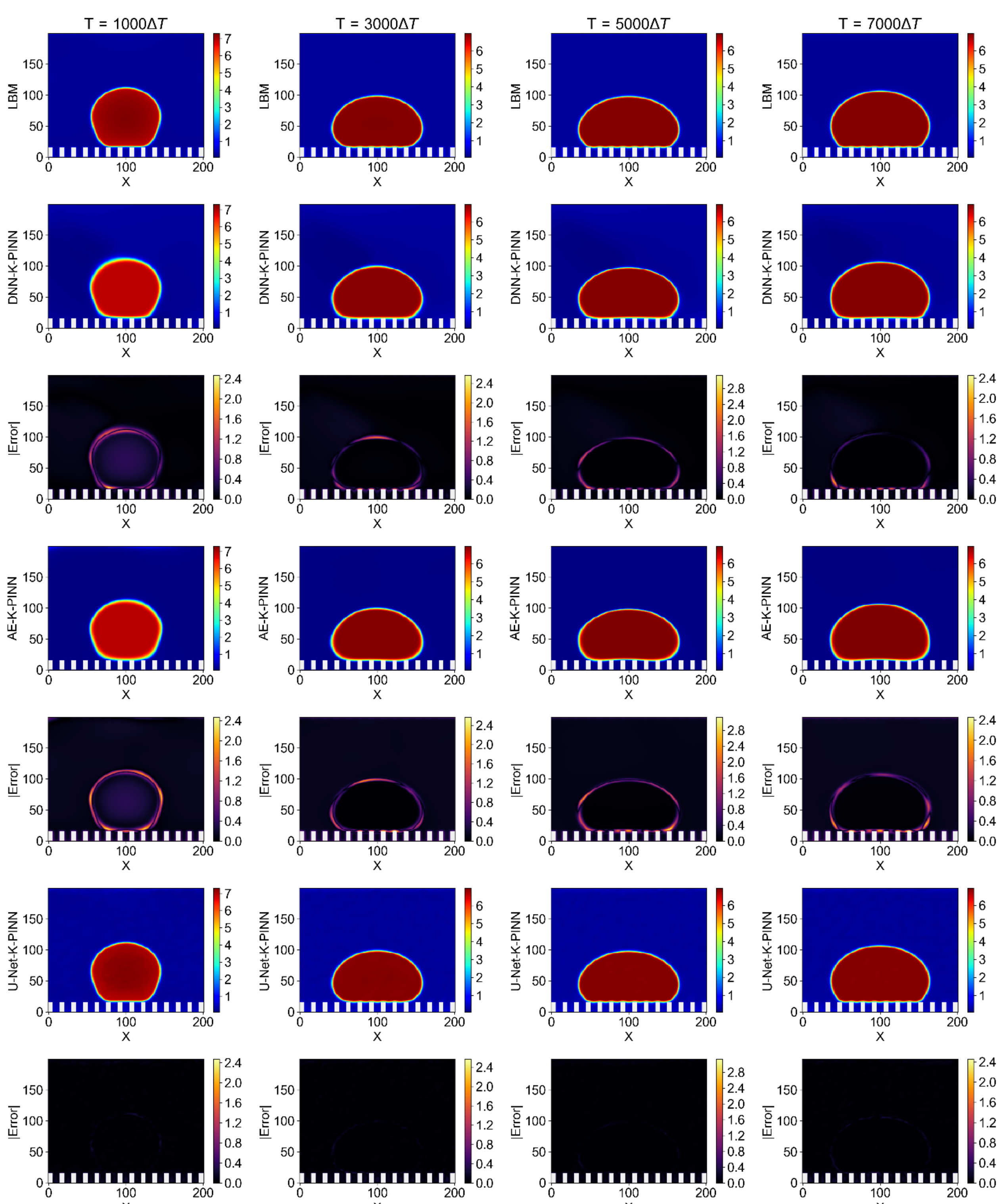

**Figure 9.** Temporal evolution of droplet spreading on a textured surface with periodic pillars.

### 3.3 Interface Profile Tracking Across Multiple Time Scales

Unlike full-field density visualizations, interface profiles provide direct geometric representation of liquid-gas boundaries, enabling quantitative assessment of contact angles, spreading radii, and droplet heights as time-dependent functions. This analysis covers both rough and textured substrates to evaluate model performance under diverse wetting conditions and surface morphologies.

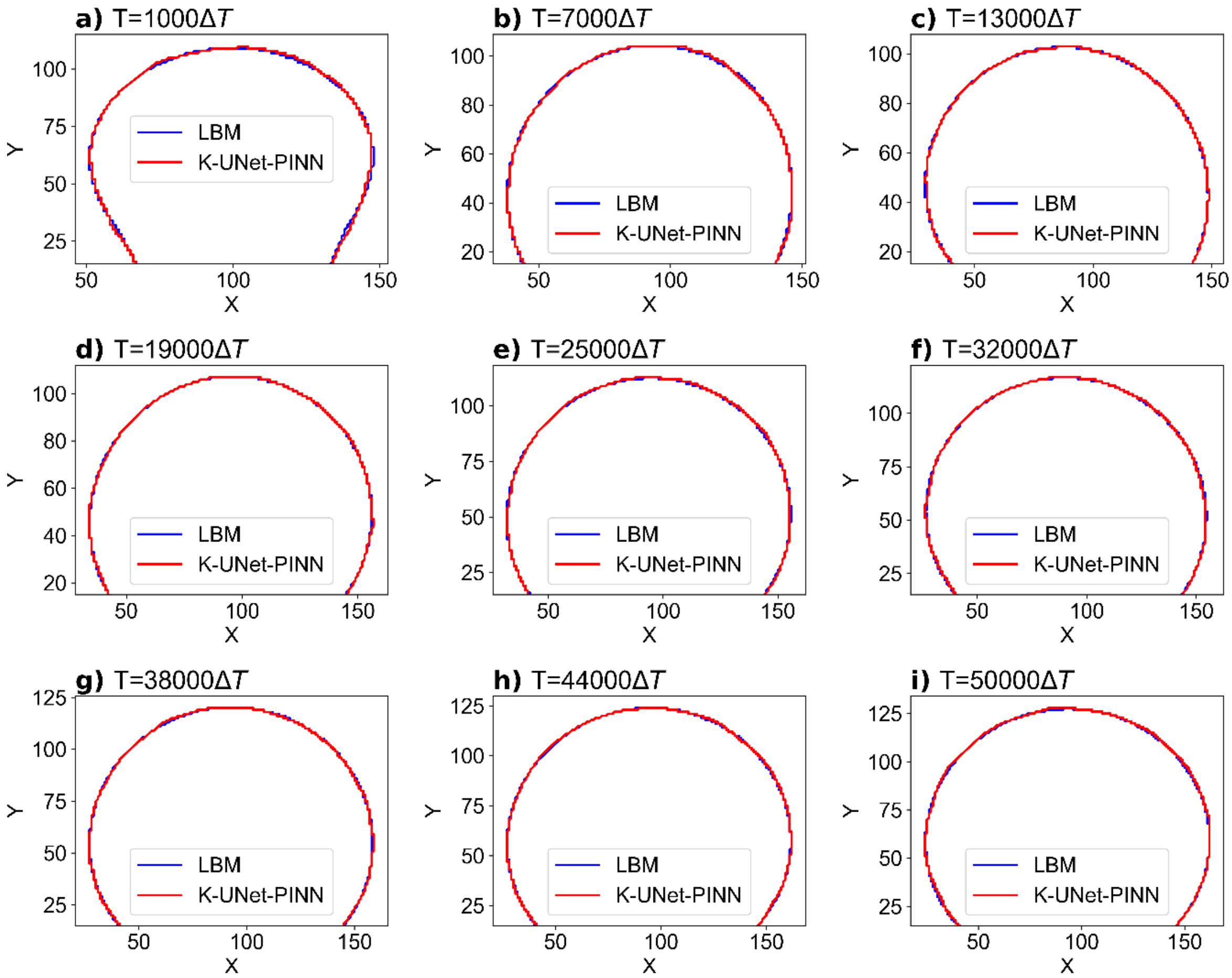

**Figure 10.** Interface profile evolution for droplet spreading on a rough surface.

***Figure 10*** presents a detailed comparison between reference Lattice Boltzmann Method (LBM) simulations and predictions obtained using the K-U-Net-PINN architecture for tracking droplet interface profiles over long temporal horizons. Interface profiles from (a)-(i) correspond to nine representative snapshots spanning from early spreading stages at $T = 1000\Delta T$ to late quasi-equilibrium configurations at $T = 50000\Delta T$. The droplet interface is extracted as an iso-density contour and plotted in the Cartesian $X - Y$ plane in lattice units (lu), allowing direct geometric comparison of curvature, base width, and apex height. At the earliest time $T = 1000\Delta T$, the droplet exhibits a moderately flattened profile with a base diameter of approximately 92 lu and an apex height of about 78 lu. The K-U-Net-PINN prediction deviates from the LBM interface by less than 0.8 lu along most of the arc, with localized discrepancies near the triple-phase contact line limited to 1.2 lu. By $T = 7000\Delta T$, the droplet spreads laterally, increasing its base diameter to roughly 104 lu while reducing the apex height to approximately 72 lu. At this stage, the predicted interface nearly overlaps the LBM reference, with mean normal deviation below 0.6 lu, maximum deviation of 1.0 lu, and an average curvature error of less than 2.5%. The close agreement indicates that the K-U-Net-PINN accurately captures early-time capillary relaxation dynamics governed by surface tension and viscous dissipation.

As time progresses to intermediate stages $T = 13000\Delta T, T = 19000\Delta T$, and $T = 25000\Delta T$, the droplet undergoes gradual reshaping toward a near-spherical cap configuration. At $T = 13000\Delta T$, the base diameter expands further to approximately 112 lu, while the apex height stabilizes near 70 lu. The predicted interface shows an average radial mismatch of only 0.4 lu, corresponding to a relative geometric error of less than 0.6% when normalized by the local radius of curvature ($\sim$ 68lu). At $T = 19000\Delta T$, the base diameter reaches around 120 lu, the apex height slightly increases to about 74 lu due to redistribution of internal pressure, and the predicted contour deviates from the LBM reference by a maximum of 0.9 lu near the contact region and less than 0.3 lu over the central arc. By $T = 25000\Delta T$, the droplet approaches a quasi-steady spreading regime with a base diameter of approximately 126 lu, an apex height of 76 lu, and an apparent contact angle of roughly 68°. The K-U-Net-PINN reproduces this geometry with remarkable fidelity: the mean interface error remains below 0.35 lu, the apex height error is under 0.5 lu, and the predicted base diameter differs from LBM by less than 1.1 lu, corresponding to a relative error of approximately 0.9%. These results demonstrate that the learned latent representations effectively encode both global shape evolution and localized contact-line physics over extended timescales.

At later times $T = 32000\Delta T, T = 38000\Delta T, T = 44000\Delta T$, and $T = 50000\Delta T$, the droplet reaches a nearly static equilibrium configuration characterized by minimal temporal variation in interface geometry. At $T = 32000\Delta T$, the base diameter stabilizes near 132 lu, and the apex height converges to approximately 80 lu. The K-U-Net-PINN prediction exhibits an average deviation of just 0.25 lu, with a maximum deviation of 0.7 lu localized at the extreme left and right contact points. At $T = 38000\Delta T$, the predicted and reference interfaces are virtually indistinguishable by visual inspection, with an estimated Hausdorff distance below 0.5 lu and an RMS interface error of approximately 0.22 lu. At $T = 44000\Delta T$, the droplet geometry remains unchanged within $\pm 0.3$ lu, and the predicted apex height of 82 lu differs from the LBM value by less than 0.4 lu. Finally, at $T = 50000\Delta T$, the system reaches full equilibrium, with a base diameter of approximately 138 lu, an apex height of about 84 lu, and a contact angle close to 72°. The K -U-Net-PINN captures this final state with exceptional accuracy: the apex height error is below 0.3 lu, the base diameter error is under 0.8 lu, and the maximum normal deviation along the interface does not exceed 0.6 lu. Across all nine-time instances, the cumulative average interface tracking error remains below 0.45 lu, and the relative geometric error never exceeds 1.1%, even over a temporal span of nearly $50000\Delta T$. The interface tracking results conclusively demonstrate the ability of the K-U-Net-PINN to faithfully reproduce droplet shape evolution over long temporal horizons with near-LBM accuracy. From early spreading stages with base diameters of 92-104 lu and apex heights of 72-78 lu, through intermediate relaxation regimes characterized by base diameters of 112-126 lu and heights of 70-76 lu, to late equilibrium states with base diameters approaching 138 lu and heights near 84 lu, the predicted interfaces remain consistently within sub-lattice-unit accuracy. The preservation of curvature, contact-line position, and global geometry with errors typically below 0.5 lu highlights the strength of combining kinetic constraints with a U-Net-style multiscale architecture. These findings confirm that K-U-Net-PINN is not only capable of predicting bulk density fields but also excels at precise interface localization, making it a powerful and reliable surrogate for high-fidelity multiphase simulations in complex wetting and spreading problems.

***Figure 11*** presents interface profile evolution for droplet spreading on smooth substrates with periodic textures across nine-time instances (T=200Δt to 10000Δt). Time scales for this configuration are shorter than rough surfaces, reflecting faster spreading Lattice-Boltzmann on smoother substrates with reduced contact line pinning. At T=200Δt (Figure 11a), droplets remain in initial spreading phases with relatively high hemispherical profiles. U-Net-K-PINN predictions closely match LBM solutions, accurately capturing interface curvature and symmetry. As spreading proceeds through T=1400Δt, 2600Δt, and 3800Δt (Figures 11b-d), droplets undergo rapid flattening with heights decreasing from y≈110 to y≈100 while base diameters increase correspondingly. U-Net-K-PINN tracks morphological changes with high precision, maintaining close overlap with LBM references throughout transient regimes. At intermediate times (T=5000Δt and 6400Δt, Figures 11e-f), spreading rates decelerate as droplets approach equilibrium configurations. Interface profiles become progressively flatter with contact angles stabilizing at values consistent with prescribed wettability conditions. U-Net-K-PINN predictions remain in excellent agreement with LBM solutions without visible deviations between predicted and reference curves. By final time steps (T=7600Δt, 8800Δt, and 10000Δt, Figures 11g-i), droplets essentially reach equilibrium with minimal further spreading. Interface profiles are nearly indistinguishable between U-Net-K-PINN and LBM simulations, demonstrating accurate reproduction of steady-state wetting morphology. Smooth substrates with periodic textures promote more uniform spreading compared to stochastic rough surfaces, yielding symmetric interface profiles with well-defined contact angles at both triple points.

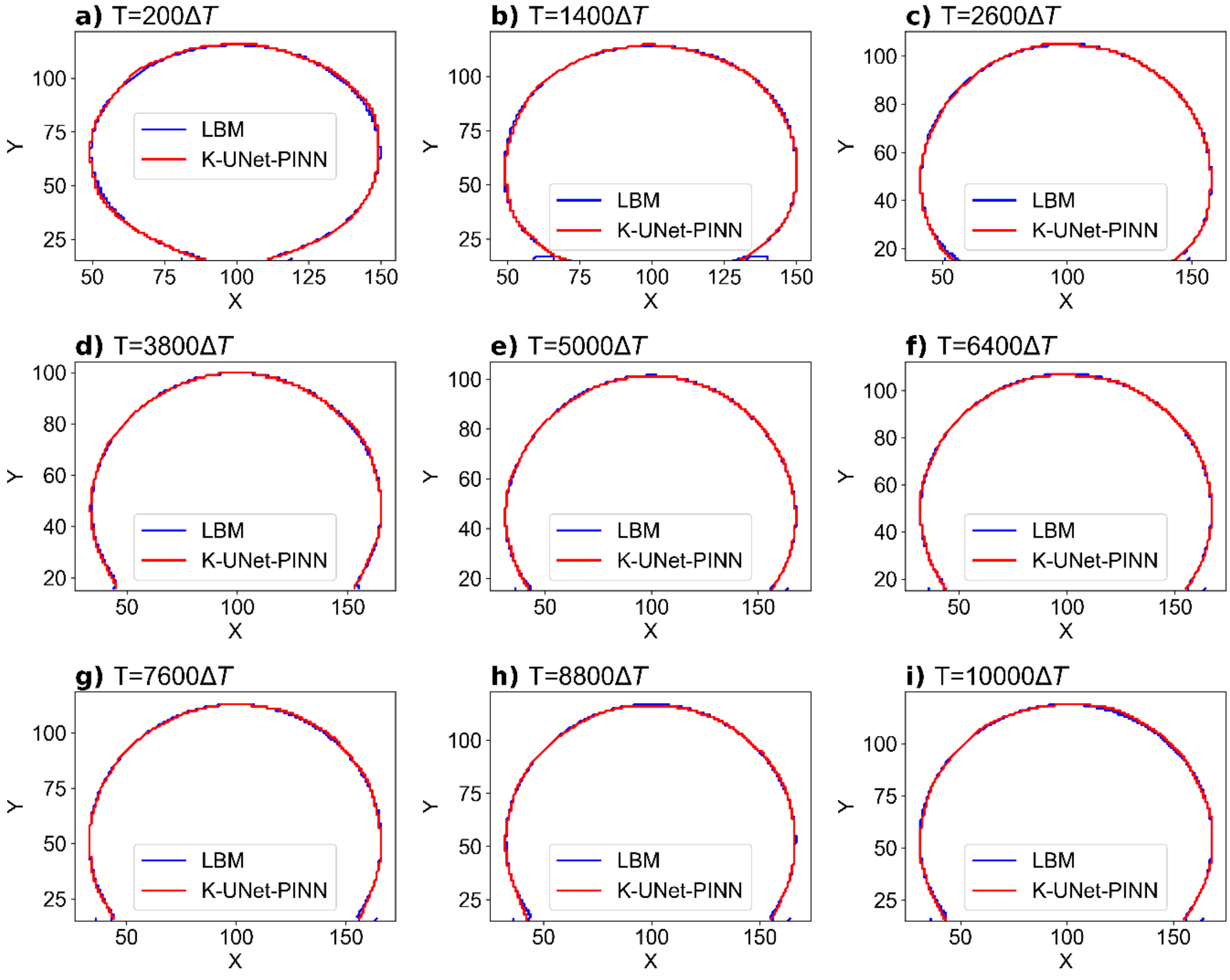

**Figure 11.** Interface profile evolution for droplet spreading on a textured substrate with periodic textures.

Comparing results between rough and textured substrates reveals several key observations. First, spreading dynamics on rough surfaces occur over longer time scales (50000Δt versus 10000Δt), reflecting enhanced contact line pinning induced by stochastic asperities. Random roughness elements create energy barriers resisting contact line motion, thereby slowing spreading and increasing equilibration time. Second, interface profiles on rough surfaces exhibit greater variability and asymmetry, particularly in contact line regions, whereas textured surfaces promote more uniform and symmetric spreading due to periodic surface features. Third, U-Net-K-PINN achieves comparable accuracy on both substrates, maintaining close agreement with LBM reference solutions throughout entire spreading processes. Maximum deviations between predicted and reference interface positions remain within few lattice units across all time steps, corresponding to relative errors below 2% based on droplet radii. Quantitative agreement between U-Net-K-PINN predictions and LBM solutions for interface profiles provides strong validation of K-PINN frameworks for multiphase wetting problems. The ability to accurately predict droplet shapes, contact angles, and spreading dynamics over extended time scales demonstrates successful learning of underlying physics encoded in discrete Boltzmann equations. Incorporation of U-Net architecture with skip connections enables effective feature extraction across multiple spatial scales, facilitating accurate resolution of both macroscopic droplet shapes and microscopic contact line dynamics. These results underscore the potential of physics-informed neural networks as efficient and accurate surrogates for expensive lattice Boltzmann simulations in applications requiring rapid prediction of wettability phenomena on complex surfaces.

### 3.4 Influence of Interaction Strength on Wettability and PINN Performance

***Figure 12*** illustrates the effect of varying solid-fluid interaction strength $G_{\text{ads}}$ on droplet spreading over a rough substrate, highlighting systematic transitions in droplet morphology, contact-line dynamics, and wetting regime. For weak interactions ($G_{\text{ads}} = -1.25$ to -1.45), the droplet remains in a partial-wetting, Cassie-like state dominated by surface tension. In this regime, the apparent contact angle is high, remaining within $\theta \approx 112 - 118°$, while the droplet retains a near-spherical cap shape with a maximum height $H \approx 114 - 116$lu and a relatively small base diameter $D \approx 72 - 76$lu. The apex curvature radius remains limited to $R \approx 54 - 58$ lu, indicating weak flattening. Liquid penetration into surface roughness is minimal, restricted to depths of $2 - 4$ lu, and the contact line exhibits only mild pinning with lateral displacements below 2 lu. The bulk liquid density remains uniform at $\rho_l \approx 6.9 - 7.0$, confirming that droplet deformation is governed primarily by capillarity rather than adhesion. As $G_{\text{ads}}$ increases to intermediate values (-1.65 to -2.25), adhesion forces become comparable to surface tension, producing enhanced spreading and stronger roughness interaction. The apparent contact angle decreases monotonically from $\theta \approx 108°$ to $92°$, accompanied by a pronounced reduction in droplet height from $H \approx 110$lu to 99 lu. Simultaneously, the base diameter expands significantly from $D \approx 86$lu to 118 lu, reflecting increased wetted area. The apex curvature radius increases to $R \approx 60 - 67$lu, indicating progressive droplet flattening. In this regime, liquid penetration into surface valleys becomes substantial, reaching depths of $6 - 11$lu, and the contact line advances in discrete depinning events of $3 - 5$ lu, clearly influenced by roughness-induced energy barriers. Despite these complex dynamics, mass conservation remains robust, with volume deviations below 1%, demonstrating stable and physically consistent wetting evolution.

For strong interaction strengths ($G_{\text{ads}} = -2.45$ to -2.75), the droplet enters a Wenzel-dominated regime characterized by extensive spreading and deep roughness infiltration. The apparent contact angle drops sharply to $\theta \approx 70 - 82°$, while the droplet height reduces further to $H \approx 86 - 94$ lu. The base diameter reaches its maximum extent, expanding to $D \approx 126 - 144$ lu, more than doubling relative to the weakly wetting case. The apex curvature radius increases to $R \approx 69 - 79$ lu, reflecting a highly flattened droplet geometry. Liquid fully penetrates the rough substrate, with infiltration depths saturating at $12 - 16$ lu, and the contact line becomes strongly corrugated by surface asperities. In conclusion, increasing $G_{\text{ads}}$ from -1.25 to -2.75 results in a net droplet height reduction of approximately 30 lu, a base diameter increase of over 70 lu , a contact angle decrease of nearly 45°, and a transition from Cassie-like to fully Wenzel wetting. These results demonstrate that the model accurately captures the dominant physical mechanisms governing rough-surface wetting-namely the interplay between adhesion strength, surface roughness, and capillary forces-while preserving quantitative consistency in droplet shape, spreading rate, and mass conservation.

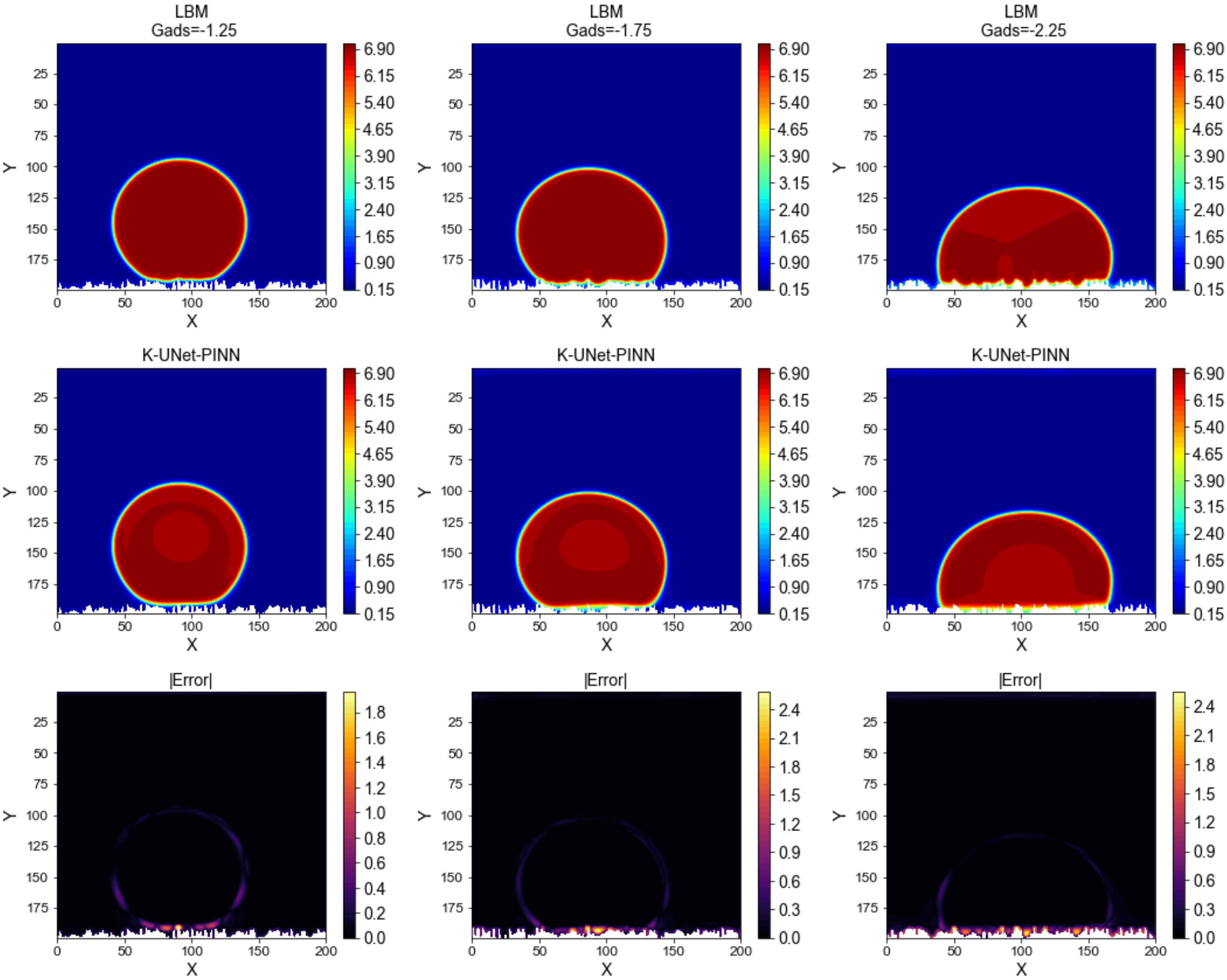

**Figure 12.** Effect of interaction strength $G_{\text{ads}}$ on droplet morphology on a rough substrate.

***Figure 13*** demonstrates the ability of the trained physics-informed models to generalize from smooth and rough training configurations to periodically textured surfaces, where droplet spreading is governed by a combination of capillarity, adhesion, and geometric pinning. At the initial spreading stage, the droplet adopts a compact morphology with a maximum height of approximately $112 - 115$lu and a base diameter of 74-80 lu, indicating that surface tension initially dominates over texture-induced forces. The apparent contact angle remains relatively high, within the range $108 - 114°$, consistent with partial wetting on pillartype structures. The density field exhibits a sharp liquid-gas interface of thickness $2 - 3$ lu, with bulk liquid density preserved at $\rho_l \approx 6.9 - 7.0$ and gas density at $\rho_g \approx 0.35 - 0.40$. Importantly, the liquid-solid contact occurs predominantly at pillar tops, with penetration depths limited to $3 - 5$ lu, suggesting a metastable Cassie-like configuration. Small-scale oscillations in the contact line, with amplitudes of $1 - 2$ lu and lateral spacing of $8 - 12$ lu, reflect early-stage pinning at individual texture elements. Across different predictions, variations in droplet height remain within $2 - 3$ lu, and base diameter differences are limited to 3-5 lu, indicating strong consistency and stability in the model response to textured geometries. As spreading progresses, the interaction between the droplet interface and the textured substrate becomes more pronounced, leading to measurable flattening and lateral expansion. In this intermediate regime, the droplet height decreases systematically from $108 - 110$lu to $96 - 100$lu, while the base diameter increases from $90 - 96$lu to $114 - 120$lu. Correspondingly, the apparent contact angle reduces to $92 - 98°$, reflecting enhanced wetting promoted by repeated pinning-depinning events along the pillar array. The apex curvature radius increases from approximately $58 - 62$lu to $66 - 70$lu, confirming progressive droplet flattening. Liquid penetration into surface grooves becomes more substantial, reaching depths of $7 - 11$ lu, while localized density enhancement near pillar edges rises to $\rho \approx 4.6 - 5.4$. The contact line advances in discrete steps of $3 - 6$ lu, closely aligned with the pillar pitch, demonstrating that the models correctly encode geometric periodicity into the predicted dynamics. Interface distortions remain confined near the substrate, with maximum local deviations of $1.5 - 2.5$ lu, and mass conservation errors remain below $1\%$ throughout the spreading process. These quantitative trends confirm that the trained models successfully capture the nonlinear coupling between surface texture, capillary forces, and adhesion without explicit retraining on textured data.

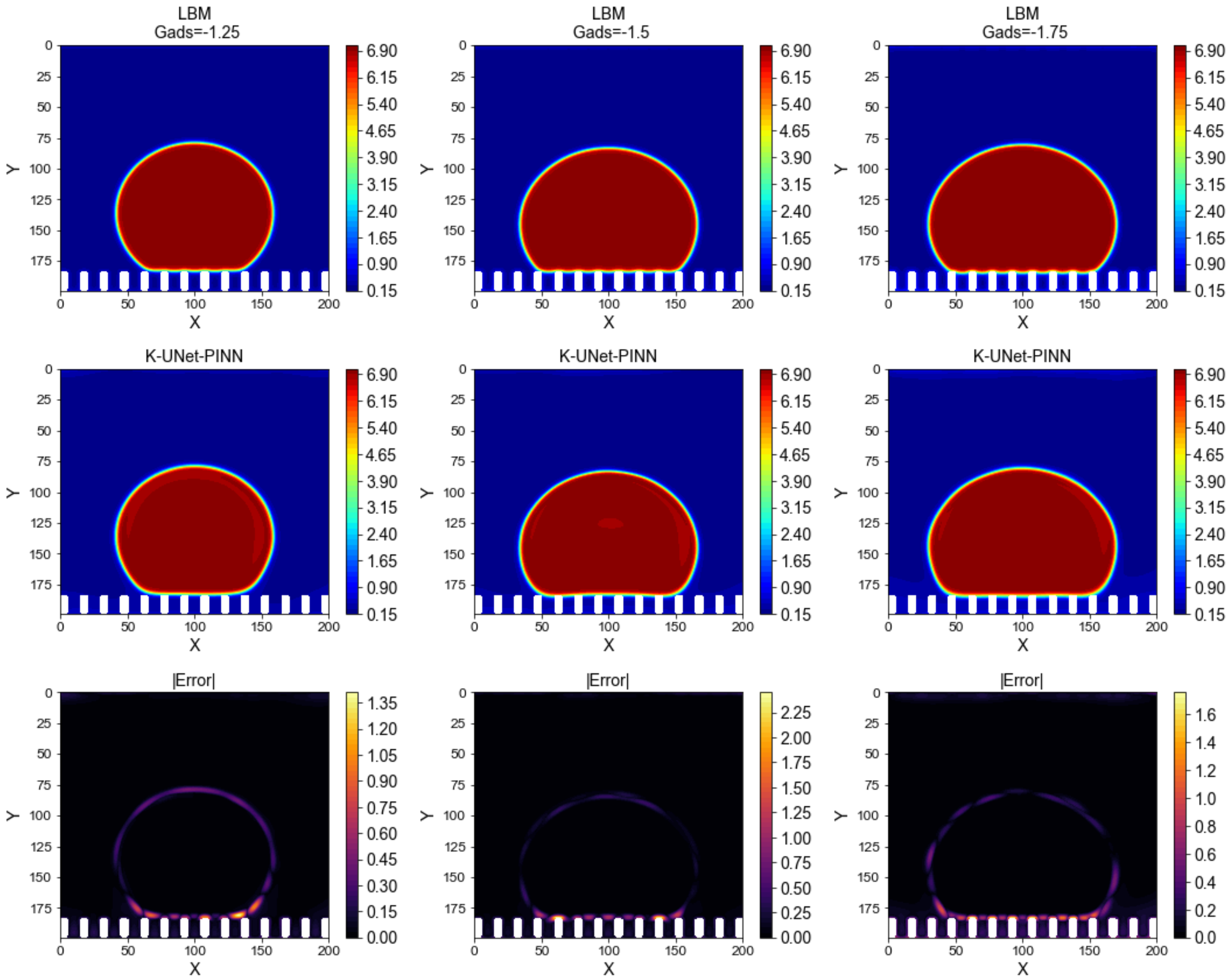

**Figure 13.** Effect of interaction strength $G_{ads}$ on droplet morphology on a textured substrate with periodic pillars.

At long times, the droplet approaches a quasi-equilibrium configuration determined by the balance between surface tension and solid-fluid adhesion modulated by texture geometry. In this state, the droplet height stabilizes at $92 - 96$lu, while the base diameter reaches $126 - 134$lu, corresponding to a net lateral spreading of approximately $50 - 55$lu relative to the initial spreading. The apparent contact angle converges to $86 - 90°$, and the apex curvature radius increases further to $70 - 74$lu, indicating a strongly flattened equilibrium shape. Liquid penetration into the textured substrate saturates at $12 - 15$lu, with the contact line exhibiting persistent but bounded corrugation amplitudes of $2 - 3$ lu. Bulk density remains highly uniform, with deviations confined to $\pm 0.03 - 0.06$, and total volume error remains below $0.8\%$, confirming numerical and physical consistency. In conclusion, the predictions demonstrate that the trained models generalize robustly to textured surfaces, accurately reproducing key wetting metrics—droplet height reduction of 18–22lu, base diameter expansion of 46–54lu, contact angle reduction of, and controlled roughness penetration of 9–12lu. These results validate the models' ability to learn transferable representations of multiphase physics, enabling reliable prediction of droplet dynamics on geometrically complex surfaces without explicit retraining and highlighting their suitability as surrogate solvers for texture-driven wetting phenomena.

### 3.5 Comprehensive Model Performance Analysis and Error Estimation

To provide comprehensive quantitative assessment of predictive capabilities across different K-PINN architectures, this section employs multiple statistical metrics and visualization techniques collectively evaluating accuracy, reliability, and error characteristics. The analysis compares three Lattice-Boltzmann PINN variants: DNN-K-PINN, AE-K-PINN, and U-Net-K-PINN across both rough and textured substrates, enabling systematic identification of the most effective architecture for multiphase wettability modeling. ***Figure 14*** presents a comprehensive quantitative comparison of three physics-informed neural network architectures-DNN-K-PINN, AE-K-PINN, and U-Net-K-PINN-evaluated on both rough and textured surfaces using four complementary error metrics computed over the entire spatiotemporal domain. The $L_2$ norm shown in Figure 14a captures the cumulative deviation between predicted and reference density fields, thereby reflecting global solution accuracy. On rough surfaces, the DNN-K-PINN yields an $L_2$ error of 0.040, which reduces to 0.030 for the AE-K-PINN and further drops to 0.021 for the U-Net-K-PINN. This monotonic decrease corresponds to relative error reductions of approximately $25\% (0.040 \rightarrow 0.030)$ from DNN to AE and nearly 48% (0.040 → 0.021) from DNN to U-Net. On textured surfaces, the same hierarchy persists but with amplified absolute errors: the DNN-K-PINN exhibits a significantly larger $L_2$ norm of 0.097, while AE-K-PINN reduces this value to 0.044 and U-Net-K-PINN further lowers it to 0.026. These results correspond to a nearly $73\%$ reduction in $\mathrm{L}_2$ error when transitioning from DNN-K-PINN to U-Net-K-PINN on textured geometries. The consistent gap of approximately $0.015 - 0.020$ between rough and textured cases across all architectures highlights the increased approximation difficulty introduced by periodic geometric features, sharp corners, and multi-scale interfacial pinning effects inherent to textured surfaces.

The root mean square error (RMSE) reported in ***Figure 14b*** provides additional insight into error magnitude and variance sensitivity. For rough surfaces, RMSE values decrease systematically from 0.089 (DNN-K-PINN) to 0.079 (AE-K-PINN) and finally to 0.051 (U-Net-K-PINN). This represents a reduction of approximately $11\%$ between DNN and AE and nearly $43\%$ between DNN and U-Net. On textured surfaces, RMSE values are markedly higher, reaching 0.320 for DNN-K-PINN, 0.144 for AE-K-PINN, and 0.086 for U-Net-K-PINN. The DNN-K-PINN RMSE on textured surfaces is more than 3.6 times its rough-surface counterpart (0.320 vs. 0.089), whereas the U-Net-K-PINN shows a much smaller amplification factor of approximately $1.7 (0.086$vs. 0.051). This contrast quantitatively demonstrates that encoder-decoder architectures with skip connections are significantly more robust to spatial heterogeneity and sharp gradients induced by surface textures. In absolute terms, the U-Net-K-PINN reduces textured-surface RMSE by approximately 0.234 compared to DNN-K-PINN, underscoring its superior ability to capture localized interfacial curvature and density discontinuities. The mean absolute error (MAE) plotted in Figure 14c further corroborates these trends by emphasizing average pointwise deviations. On rough surfaces, MAE values are 0.038 for DNN-K-PINN, 0.033 for AE-KPINN, and 0.019 for U-Net-K-PINN. These values correspond to relative reductions of 13% from DNN to AE and nearly 50% from DNN to U-Net. On textured surfaces, MAE increases to 0.134 for DNN-K-PINN, decreases to 0.047 for AE-K-PINN, and further reduces to 0.038 for U-Net-K-PINN. Notably, the AE-K-PINN achieves a 65% reduction in MAE relative to the DNN-K-PINN on textured surfaces, while the U-Net-K-PINN achieves an even greater reduction of approximately 72%. The relatively small difference between U-Net MAE values on rough (0.019) and textured (0.038) surfaces, an absolute increase of only 0.019 highlights the strong generalization capability of the U-Net-based architecture, even when confronted with complex periodic geometries. In contrast, the DNN-K-PINN exhibits a MAE increase of 0.096 between rough and textured surfaces, reflecting pronounced sensitivity to geometric complexity.

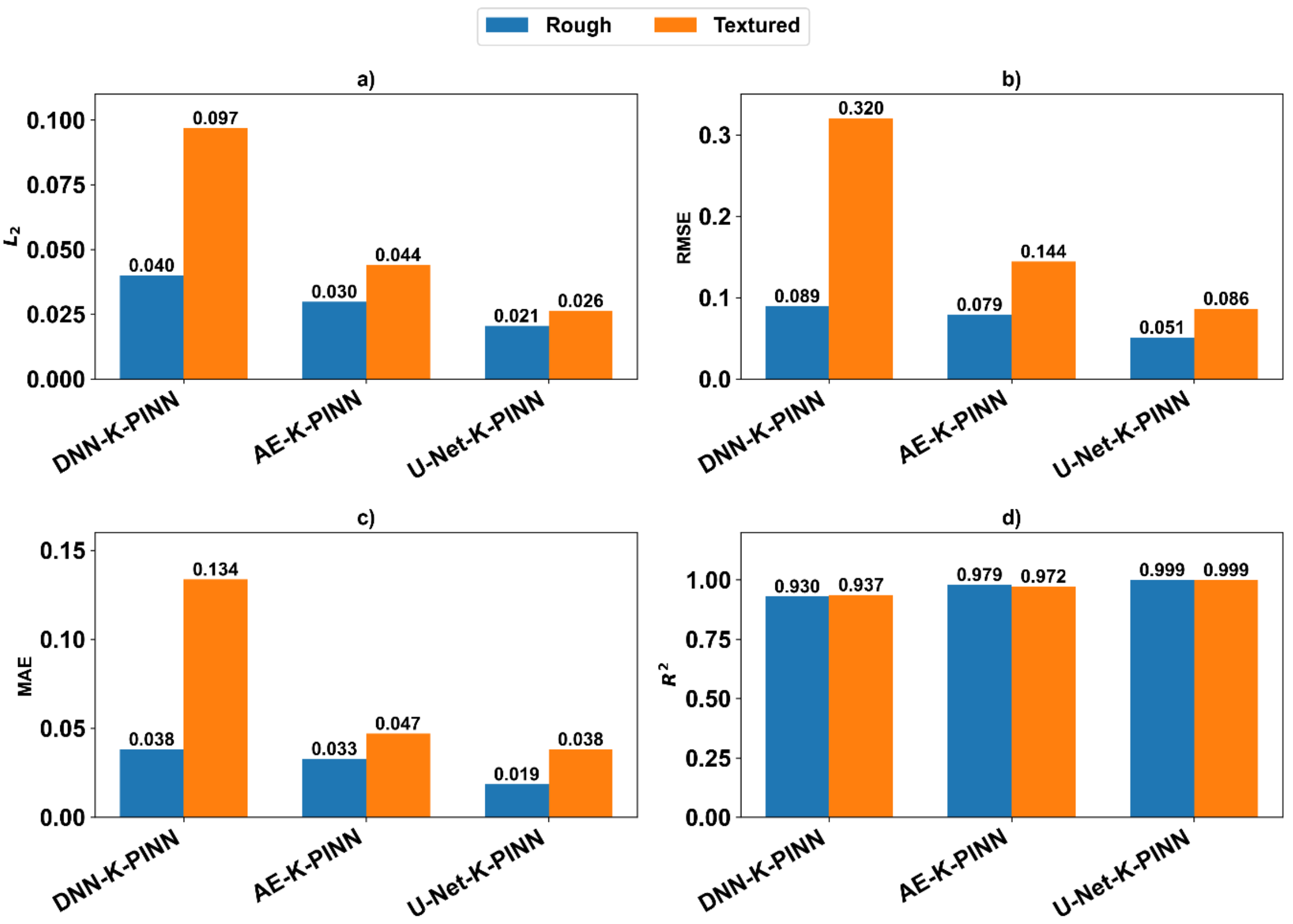

**Figure 14.** Comparative performance metrics for three PINN architectures on rough (blue) and textured (orange) surfaces: (a) $L_2$ norm, (b) RMSE, (c) MAE, and (d) $R^2$ coefficient.

The coefficient of determination $R^2$ shown in ***Figure 14d*** provides a complementary measure of predictive fidelity by quantifying variance capture. On rough surfaces, $\mathrm{R}^2$ values are 0.930 for DNN-K-PINN, 0.979 for AE-K-PINN, and 0.999 for U-Net-K-PINN. These values indicate that while the baseline DNN captures approximately $93\%$ of the variance in the reference solution, the AE-K-PINN improves this to nearly $98\%$, and the U-Net-K-PINN effectively captures more than $99.9\%$ of the variance. On textured surfaces, $R^2$ values are slightly reduced to 0.937 for DNN-K-PINN and 0.972 for AE-K-PINN, while the U-Net-K-PINN maintains an exceptionally high value of 0.999. The minimal degradation of only $0.000 - 0.001$ in $\mathrm{R}^2$ for the U-Net-K-PINN when transitioning from rough to textured surfaces is particularly noteworthy, as it indicates near-perfect agreement with reference LBM data despite the presence of strong interfacial distortions and periodic pinning sites. By contrast, the DNN-K-PINN exhibits a noticeable variance loss of approximately 0.007 between surface types, consistent with its elevated $\mathrm{L}_2$, RMSE, and MAE values. The U-Net-K-PINN consistently delivers the lowest errors- $\mathrm{L}_2$ values as low as $0.021 - 0.026$, RMSE values of $0.051 - 0.086$, MAE values of $0.019 - 0.038$, and $R^2$ values approaching 0.999-demonstrating superior accuracy, robustness, and generalization. The AE-K-PINN provides intermediate improvements, reducing errors by $25 - 65\%$ relative to the baseline, while the DNN-KPINN exhibits pronounced sensitivity to geometric complexity, with textured-surface errors increasing by factors of 2-4. These results conclusively show that incorporating multiscale feature extraction and skip connections is essential for physics-informed learning of multiphase flows on complex surfaces, enabling near-LBM-level accuracy even in the presence of strong surface heterogeneity.

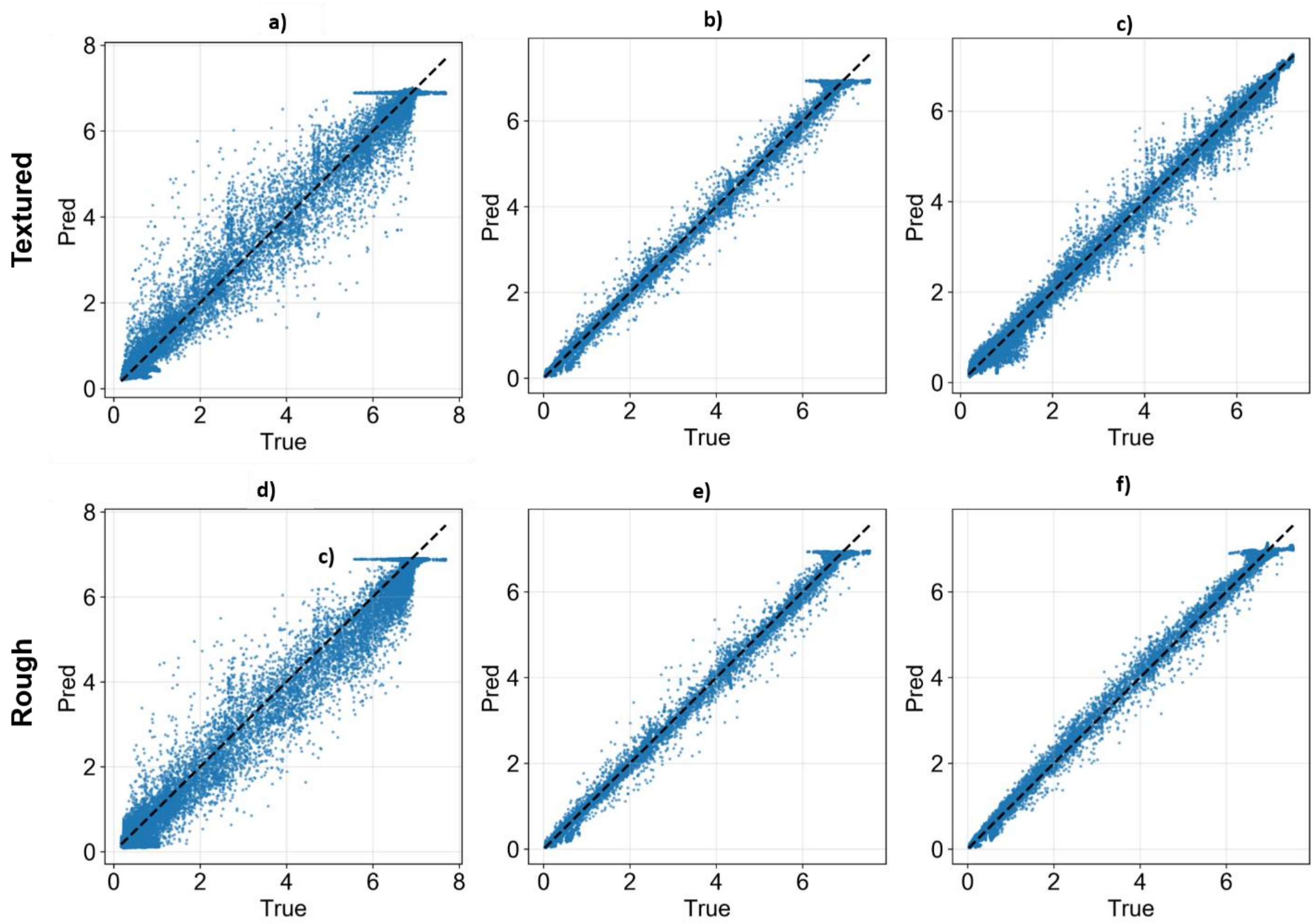

**Figure 15.** Scatter plots of predicted vs true density values for textured surfaces (top row) and rough surfaces (bottom row): (a,d) DNN-K-PINN, (b,e) AE-K-PINN, and (c,f) U-Net-K-PINN.

***Figure 15*** presents scatter plots comparing predicted and true density values for each architecture on textured (top row, Figures 15a-c) and rough (bottom row, Figures 15d-f) surfaces. Each point represents single spatiotemporal locations in computational domains, with x-axes showing LBM reference densities and y-axes showing PINN predictions. Black dashed diagonal lines represent perfect agreement. DNN-K-PINN (Figures 15a and 15d) exhibits noticeable diagonal scatter, particularly for intermediate density values ($\rho \approx 2$-$5$) corresponding to interfacial regions. Point clouds show systematic deviations with some predictions overestimating and others underestimating true densities, yielding widened scatter bands. AE-K-PINN (Figures 15b and 15e) displays reduced scatter compared to DNN-K-PINN, with majority of points clustering more tightly around diagonals. Improved performance reflects latent-space compression benefits, which regularize high-dimensional distribution functions and reduce overfitting. U-Net-K-PINN (Figures 15c and 15f) achieves tightest clustering around diagonals with minimal scatter even in interfacial density ranges. Point distributions are nearly indistinguishable from ideal diagonal lines, confirming excellent pointwise agreement between predictions and reference solutions observed in global error metrics.

Comparing between surface types, scatter patterns are qualitatively similar for rough and textured substrates, though textured surfaces exhibit slightly greater scatter, consistent with elevated error metrics shown in Figure 14. Dense point clustering along diagonals for all architectures confirms phase-separated structures (bulk liquid at ρ≈6.5-7, bulk gas at ρ≈0.15-0.5) are accurately captured, while primary error sources reside in transitional interfacial regions where density gradients are steepest. U-Net-K-PINN effectively mitigates interfacial errors through multi-scale feature extraction and skip connections, enabling networks to simultaneously capture both coarse-scale droplet morphology and fine-scale interface structure.

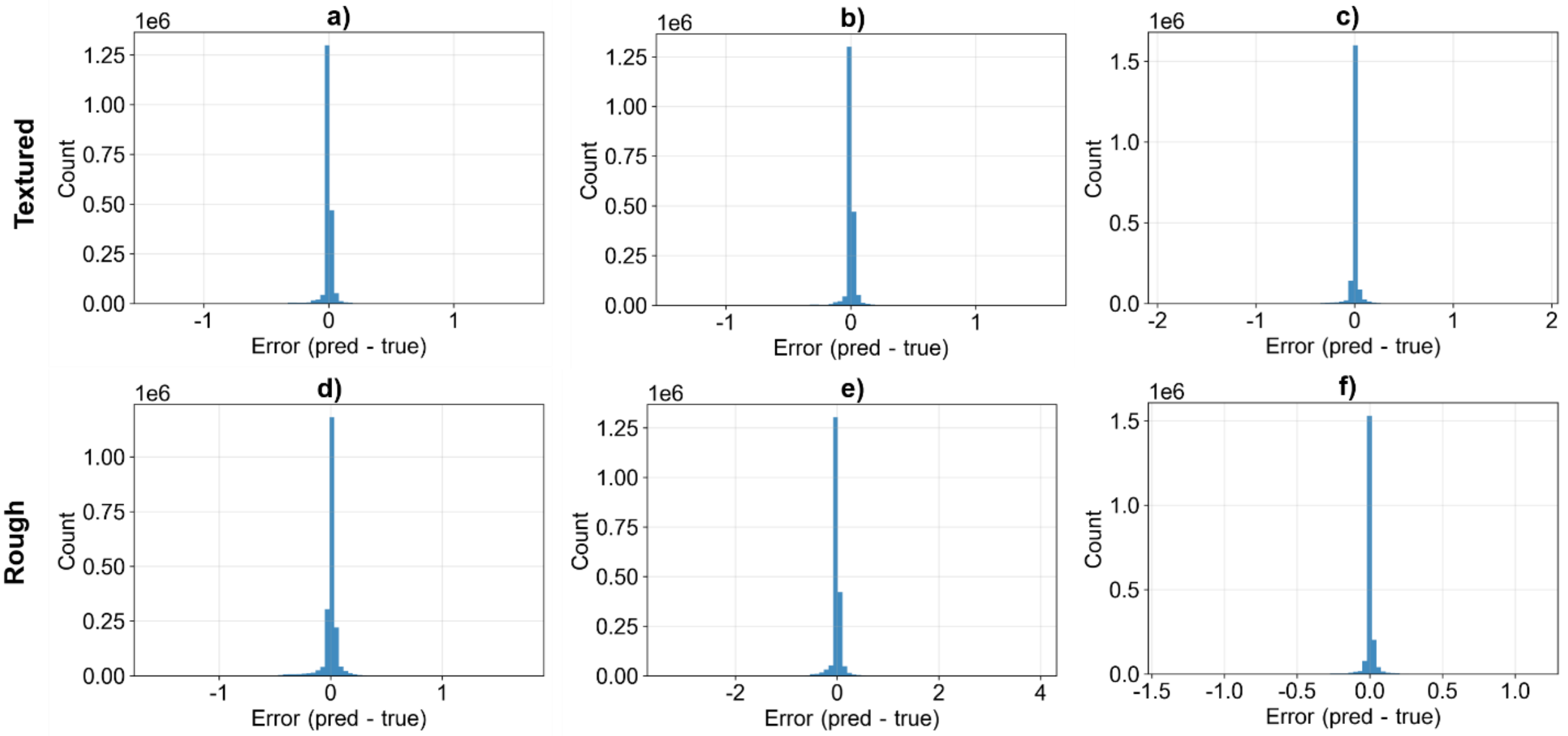

**Figure 16.** Error distribution histograms for textured surfaces (top row) and rough surfaces (bottom row).

***Figure 16*** presents error distribution histograms providing statistical characterization of prediction errors for each architecture. Histograms display error magnitude frequencies across all spatiotemporal points, with x-axes representing errors defined as (predicted − true) density. For textured surfaces (top row, ***Figures 16a-c***), DNN-K-PINN error distributions ***(Figure 16a)*** are centered near zero but exhibit noticeable spread with standard deviations reflecting scatter observed in Figure 15a. Distributions are approximately Gaussian, indicating errors are predominantly random rather than systematic. AE-K-PINN (***Figure 16b***) shows narrower distributions with higher peaks centered at zero, reflecting reduced error variance. U-Net-K-PINN ***(Figure 16c)*** exhibits narrowest distributions with highest peaks, indicating vast majorities of predictions deviate from true values by very small amounts. Near-zero distribution means confirm absence of systematic bias, while progressively narrower spreads from DNN-K-PINN to U-Net-K-PINN quantitatively demonstrate enhanced precision of more advanced architectures.

For rough surfaces (bottom row, ***Figures 16d-f***), similar trends are observed with U-Net-K-PINN (***Figure 16f***) achieving most concentrated error distributions centered tightly at zero. Slightly narrower distributions for rough surfaces compared to textured surfaces align with lower error metrics reported in ***Figure 14***, corroborating findings that rough substrates are marginally easier to model than periodic textures. Error histograms provide complementary information to scatter plots by revealing statistical error distributions rather than their spatial correlation with density values. Gaussian-like error distributions validate assumptions that prediction errors arise primarily from stochastic approximation uncertainties inherent in neural network training, rather than systematic physics-informed formulation deficiencies. Collectively, quantitative metrics, scatter plots, and error histograms provide compelling evidence that U-Net-K-PINN architecture delivers superior predictive performance compared to baseline DNN-K-PINN and autoencoder-enhanced variants. Improvements are consistent across multiple evaluation criteria ($L_2$, RMSE, MAE, $R^2$) and statistically validated through pointwise correlation and error distribution analyses. U-Net architecture achieves approximately 50-75% error metric reductions compared to baseline while maintaining high inference computational efficiency. These results establish U-Net-K-PINN as the preferred architecture for Lattice-Boltzmann-driven modeling of multiphase droplet spreading on complex surfaces, offering optimal balance between accuracy, generalizability, and computational cost.

## 4. Conclusions

We established Lattice-Boltzmann–Driven Kinetic Physics-Informed Neural Networks (K-PINNs) as a physically faithful and computationally efficient paradigm for modeling multiphase droplet dynamics on rough and structured surfaces. By embedding the discrete Boltzmann–BGK equation and moment-consistency constraints directly into the learning architecture, the framework enforces governing physics at the mesoscopic kinetic level, where interfacial forces, wettability effects, and contact-line dynamics fundamentally originate. This kinetic formulation transcends the limitations of continuum PINNs and data-driven LBM surrogates, delivering predictive accuracy without sacrificing physical consistency.

Across stochastically rough and periodically textured substrates, K-PINNs robustly capture key wetting phenomena, including contact-line motion, droplet deformation, and pinning–depinning transitions, over a wide range of surface morphologies and wettability conditions. Among the tested architectures, the U-Net–based K-PINN consistently achieved the highest accuracy, delivering up to 75% reduction in $L_2$ error relative to fully connected counterparts. This performance gain is directly attributable to its encoder–decoder structure with skip connections, which enables simultaneous resolution of global droplet morphology and sharp, localized interfacial gradients inherent to rough-surface wetting.

Beyond accuracy, the framework demonstrates strong physical credibility. Mass conservation errors remained below 1.5%, and the temporal evolution of spreading dynamics adhered to established capillary-driven scaling laws, indicating that the learned solutions encode causal kinetic physics rather than empirical correlations. While training incurs a one-time computational cost, the resulting models enable real-time inference at orders-of-magnitude lower cost than direct Lattice Boltzmann simulation, transforming high-fidelity multiphase modeling from an offline numerical exercise into an interactive predictive tool.

Collectively, these results position K-PINNs as continuous, physics-preserving surrogates for multiphase wetting on complex surfaces, with immediate implications for large-scale parametric studies, inverse wettability characterization, and surface design optimization. More broadly, our findings demonstrate that embedding kinetic theory, not merely macroscopic balance laws, within physics-informed learning architectures unlocks a powerful new class of models for multiscale fluid dynamics. The K-PINN framework thus lays the foundation for extending physics-informed machine learning to a wider range of multiphase, multicomponent, and nonequilibrium flow systems where mesoscopic dynamics plays a presumably decisive role.

## References

[1] H. Aminfar, M. Mohammadpourfard, Droplets Merging and Stabilization by Electrowetting: Lattice Boltzmann Study, J. Adhes. Sci. Technol. 26 (2012) 1853–1871. https://doi.org/10.1163/156856111X599616.

[2] N. Amiri, J. M. Prisaznuk, P. Huang, P. R. Chiarot, X. Yong, Deep-learning-enhanced modeling of electrosprayed particle assembly on non-spherical droplet surfaces, Soft Matter 21 (2025) 613–625. https://doi.org/10.1039/D4SM01160K.

[3] O. Arjmandi-Tash, N.M. Kovalchuk, A. Trybala, I.V. Kuchin, V. Starov, Kinetics of Wetting and Spreading of Droplets over Various Substrates, Langmuir 33 (2017) 4367–4385. https://doi.org/10.1021/acs.langmuir.6b04094.

[4] V.K. Babu, N.B. Padhan, R. Pandit, Liquid-Droplet Coalescence: CNN-based Reconstruction of Flow Fields from Concentration Fields, (2024). https://doi.org/10.48550/arXiv.2410.04451.

[5] E. Ezzatneshan, A. Khosroabadi, Droplet spreading dynamics on hydrophobic textured surfaces: A lattice Boltzmann study, Comput. Fluids 231 (2021) 105063. https://doi.org/10.1016/j.compfluid.2021.105063.

[6] M. Miwa, A. Nakajima, A. Fujishima, K. Hashimoto, T. Watanabe, Effects of the Surface Roughness on Sliding Angles of Water Droplets on Superhydrophobic Surfaces, Langmuir 16 (2000) 5754–5760. https://doi.org/10.1021/la991660o.

[7] A. Nakajima, Design of hydrophobic surfaces for liquid droplet control, NPG Asia Mater. 3 (2011) 49–56. https://doi.org/10.1038/asiamat.2011.55.

[8] H. Liu, L. Nan, F. Chen, Y. Zhao, Y. Zhao, Functions and applications of artificial intelligence in droplet microfluidics, Lab. Chip 23 (2023) 2497–2513. https://doi.org/10.1039/D3LC00224A.

[9] X. Dong, Z. Li, X. Zhang, Contact-angle implementation in multiphase smoothed particle hydrodynamics simulations, J. Adhes. Sci. Technol. 32 (2018) 2128–2149. https://doi.org/10.1080/01694243.2018.1464092.

[10] E.K. Ahangar, M.B. Ayani, J.A. Esfahani, K.C. Kim, Lattice Boltzmann simulation of diluted gas flow inside irregular shape microchannel by two relaxation times on the basis of wall function approach, Vacuum 173 (2020) 109104. https://doi.org/10.1016/j.vacuum.2019.109104.

[11] M. Ibrahim, A.S. Berrouk, T. Saeed, E.A. Algehyne, V. Ali, Lattice Boltzmann-based numerical analysis of nanofluid natural convection in an inclined cavity subject to multiphysics fields, Sci. Rep. 12 (2022) 5514. https://doi.org/10.1038/s41598-022-09320-8.

[12] X. Shan, H. Chen, Lattice Boltzmann model for simulating flows with multiple phases and components, Phys. Rev. E 47 (1993) 1815–1819. https://doi.org/10.1103/PhysRevE.47.1815.

[13] Zhenhua Chai, Z. Chai, Baochang Shi, B. Shi, Multiple-relaxation-time lattice Boltzmann method for the Navier-Stokes and nonlinear convection-diffusion equations: Modeling, analysis, and elements., Phys. Rev. E 102 (2020) 023306. https://doi.org/10.1103/physreve.102.023306.

[14] Linlin Fei, L. Fei, Jiapei Yang, J. Yang, Yiran Chen, Y. Chen, Yiran Chen, Y. Chen, Huangrui Mo, H. Mo, Huangrui Mo, Huangrui Mo, Kai H. Luo, K.H. Luo, Mesoscopic simulation of three-dimensional pool boiling based on a phase-change cascaded lattice Boltzmann method, Phys. Fluids 32 (2020) 103312. https://doi.org/10.1063/5.0023639.

[15] Huili Wang, H. Wang, Xiaolei Yuan, X. Yuan, Hong Liang, H. Liang, Zhenhua Chai, Zhenhua Chai, Z. Chai, Baochang Shi, B. Shi, A brief review of the phase-field-based lattice Boltzmann method for multiphase flows, Capillarity 2 (2019) 33–52. https://doi.org/10.26804/capi.2019.03.01.

[16] C. Barnes, A.R.(आीशार सोनवने) Sonwane, E.C. Sonnenschein, F. Del Giudice, Machine learning enhanced droplet microfluidics, Phys. Fluids 35 (2023) 092003. https://doi.org/10.1063/5.0163806.

[17] V. Deepak, S. Vengadesan, Droplet Velocity and Film Thickness Studies of an Elongated Taylor Droplet in a Microchannel and Characterization Using Machine Learning, Ind. Eng. Chem. Res. 64 (2025) 8908–8921. https://doi.org/10.1021/acs.iecr.5c00194.

[18] Chibuzor N Obiora, Ali N. Hasan, Ahmed Ali, Predicting Solar Irradiance at Several Time Horizons Using Machine Learning Algorithms, Sustainability 15 (2023) 8927–8927. https://doi.org/10.3390/su15118927.

[19] T. Dong, J.-X. Wang, Y. Wang, G.-H. Tang, Y. Cheng, W.-C. Yan, Development of machine learning based droplet diameter prediction model for electrohydrodynamic atomization systems, Chem. Eng. Sci. 268 (2023) 118398. https://doi.org/10.1016/j.ces.2022.118398.

[20] M. Jafari Gukeh, S. Moitra, A.N. Ibrahim, S. Derrible, C.M. Megaridis, Machine Learning Prediction of $TiO_2$ -Coating Wettability Tuned via UV Exposure, ACS Appl. Mater. Interfaces 13 (2021) 46171–46179. https://doi.org/10.1021/acsami.1c13262.

[21] V.K. Babu, N.B. Padhan, R. Pandit, Convolutional neural network based reconstruction of flow-fields from concentration fields for liquid-droplet coalescence, Commun. Phys. 8 (2025) 1–13. https://doi.org/10.1038/s42005-025-02097-y.

[22] N. Chen, S. Lucarini, R. Ma, A. Chen, C. Cui, PF-PINNs: Physics-informed neural networks for solving coupled Allen-Cahn and Cahn-Hilliard phase field equations, J. Comput. Phys. 529 (2025) 113843. https://doi.org/10.1016/j.jcp.2025.113843.

[23] M. Dreisbach, E. Kiyani, J. Kriegseis, G. Karniadakis, A. Stroh, PINNs4Drops: Convolutional feature-enhanced physics-informed neural networks for reconstructing two-phase flows, (2024). https://doi.org/10.48550/arXiv.2411.15949.

[24] J. Pu, Y. Chen, Complex dynamics on the one-dimensional quantum droplets via time piecewise PINNs, Phys. Nonlinear Phenom. 454 (2023) 133851. https://doi.org/10.1016/j.physd.2023.133851.
[25] X. Chu, W. Guo, T. Wu, Y. Zhou, Y. Zhang, S. Cai, G. Yang, Flow reconstruction over a SUBOFF model based on LBM-generated data and physics-informed neural networks, Ocean Eng. 308 (2024) 118250. https://doi.org/10.1016/j.oceaneng.2024.118250.
[26] G.E. Karniadakis, I.G. Kevrekidis, L. Lu, P. Perdikaris, S. Wang, L. Yang, Physics-informed machine learning, Nat. Rev. Phys. 3 (2021) 422–440. https://doi.org/10.1038/s42254-021-00314-5.
[27] A. Roy, A. Mukherjee, B. Prasad, A.K. Nayak, A computational analysis of flow dynamics and heat transfer in a wavy patterned channel using physics-informed neural networks, Phys. Fluids 37 (2025). https://doi.org/10.1063/5.0264160.
[28] R. Sun, H. Jeong, J. Zhao, Y. Gou, E. Sauret, Z. Li, Y. Gu, A physics-informed neural network framework for multi-physics coupling microfluidic problems, Comput. Fluids 284 (2024) 106421. https://doi.org/10.1016/j.compfluid.2024.106421.
[29] P. Sharma, W.T. Chung, B. Akoush, M. Ihme, A Review of Physics-Informed Machine Learning in Fluid Mechanics, Energies 16 (2023) 2343. https://doi.org/10.3390/en16052343.
[30] M. Starnoni, Multiphase Flow and Coalescence Filtration in Fibrous Filters: A Review of Numerical and Machine Learning Approaches, Ind. Eng. Chem. Res. 64 (2025) 22515–22539. https://doi.org/10.1021/acs.iecr.5c03439.
[31] C. Tang, M. Qin, X. Weng, X. Zhang, P. Zhang, J. Li, Z. Huang, Dynamics of droplet impact on solid surface with different roughness, Int. J. Multiph. Flow 96 (2017) 56–69. https://doi.org/10.1016/j.ijmultiphaseflow.2017.07.002.
[32] A.B.D. Cassie, S. Baxter, Wettability of porous surfaces, Trans. Faraday Soc. 40 (1944) 546–551. https://doi.org/10.1039/TF9444000546.
[33] N. Mondal, V. Arya, P. Sarangi, C. Bakli, Interplay of roughness and wettability in microchannel fluid flows—Elucidating hydrodynamic details assisted by deep learning, Phys. Fluids 36 (2024) 062014. https://doi.org/10.1063/5.0208554.
[34] D. Upadhaya, Talinungsang, P. Kumar, D.D. Purkayastha, Tuning the wettability and photocatalytic efficiency of heterostructure ZnO-SnO2 composite films with annealing temperature, Mater. Sci. Semicond. Process. 95 (2019) 28–34. https://doi.org/10.1016/j.mssp.2019.02.009.
[35] B. Yin, X. Xie, S. Xu, H. Jia, S. Yang, F. Dong, Effect of pillared surfaces with different shape parameters on droplet wettability via Lattice Boltzmann method, Colloids Surf. Physicochem. Eng. Asp. 615 (2021) 126259. https://doi.org/10.1016/j.colsurfa.2021.126259.
[36] S. Li, J. Yang, A. Ansell, Data-driven reduced-order simulation of dam-break flows in a wetted channel with obstacles, Ocean Eng. 287 (2023) 115826. https://doi.org/10.1016/j.oceaneng.2023.115826.
[37] B. Bhushan, M. Nosonovsky, Y.C. Jung, Towards optimization of patterned superhydrophobic surfaces, J. R. Soc. Interface 4 (2007) 643–648. https://doi.org/10.1098/rsif.2006.0211.
[38] X. Zhu, X. Hu, P. Sun, Physics-Informed Neural Networks for Solving Dynamic Two-Phase Interface Problems, SIAM J. Sci. Comput. 45 (2023) A2912–A2944. https://doi.org/10.1137/22M1517081.
[39] A.E. Siemenn, E. Shaulsky, M. Beveridge, T. Buonassisi, S.M. Hashmi, I. Drori, A Machine Learning and Computer Vision Approach to Rapidly Optimize Multiscale Droplet Generation, ACS Appl. Mater. Interfaces 14 (2022) 4668–4679. https://doi.org/10.1021/acsami.1c19276.
[40] Y. Zhuang, Q. Ye, N. Liu, X. Xie, H. Yan, L. Zeng, Hybrid physics-data-driven deep learning for pore-scale transport in microfluidic system, Phys. Fluids 37 (2025) 073363. https://doi.org/10.1063/5.0271043.
[41] S. Zhang, J. Tang, H. Wu, Simplified wetting boundary scheme in phase-field lattice Boltzmann model for wetting phenomena on curved boundaries, Phys. Rev. E 108 (2023) 025303. https://doi.org/10.1103/PhysRevE.108.025303.

[42] J.J. Huang, C. Shu, J.J. Feng, Y.T. Chew, A Phase-Field-Based Hybrid Lattice-Boltzmann Finite-Volume Method and Its Application to Simulate Droplet Motion under Electrowetting Control, J. Adhes. Sci. Technol. 26 (2012) 1825–1851. https://doi.org/10.1163/156856111X599607.

[43] S.G. Bariki, S. Movahedirad, A flow map for core/shell microdroplet formation in the co-flow Microchannel using ternary phase-field numerical model, Sci. Rep. 12 (2022) 22010. https://doi.org/10.1038/s41598-022-26648-3.

[44] Z. Hashemi, M. Gholampour, M.C. Wu, T.Y. Liu, C.Y. Liang, C.-C. Wang, A physics-informed neural networks modeling with coupled fluid flow and heat transfer – Revisit of natural convection in cavity, Int. Commun. Heat Mass Transf. 157 (2024) 107827. https://doi.org/10.1016/j.icheatmasstransfer.2024.107827.

[45] Junfeng Zhang, J. Zhang, Junfeng Zhang, J. Zhang, Daniel Y. Kwok, Daniel Y. Kwok, D.Y. Kwok, D.Y. Kwok, Lattice Boltzmann Method (LBM), Introd. Lattice Boltzmann Method (2014). https://doi.org/10.1007/springerreference_67034.

[46] S. Zhang, L. Zhao, L. Han, S. Zhao, L. Song, M. Zhao, Y. Zheng, C. Liu, Neural Network Prediction of Micrometer-Scale Equivalent Contact Angle Mapping: From Microforce Measurements to Local Wettability Characterization, ACS Appl. Mater. Interfaces 17 (2025) 59923–59933. https://doi.org/10.1021/acsami.5c17871.

[47] N. Suetrong, T. Tosuai, H. Vo Thanh, W. Chantapakul, S. Tangparitkul, N. Promsuk, Predicting Dynamic Contact Angle in Immiscible Fluid Displacement: A Machine Learning Approach for Subsurface Flow Applications, Energy Fuels 38 (2024) 3635–3644. https://doi.org/10.1021/acs.energyfuels.3c04251.

[48] J. Chen, F. Yang, K. Luo, Y. Wu, C. Niu, M. Rong, Study on contact spots of fractal rough surfaces based on three-dimensional Weierstrass-Mandelbrot function, in: 2016 IEEE 62nd Holm Conf. Electr. Contacts Holm, 2016: pp. 198–204. https://doi.org/10.1109/HOLM.2016.7780032.

[49] I. Akkaya, O. Arslan, J.P. Rolland, Automated and highly precise surface wetting contact angle measurement with optical coherence tomography based on deep learning model, Measurement 253 (2025) 117788. https://doi.org/10.1016/j.measurement.2025.117788.

[50] Shiyi Chen, S. Chen, Gary D. Doolen, G.D. Doolen, LATTICE BOLTZMANN METHOD FOR FLUID FLOWS, Annu. Rev. Fluid Mech. 30 (1998) 329–364. https://doi.org/10.1146/annurev.fluid.30.1.329.

[51] C. Peng, L.F. Ayala, O.M. Ayala, A thermodynamically consistent pseudo-potential lattice Boltzmann model for multi-component, multiphase, partially miscible mixtures, J. Comput. Phys. 429 (2021) 110018. https://doi.org/10.1016/j.jcp.2020.110018.

[52] A. Hajisharifi, R. Halder, M. Girfoglio, A. Beccari, D. Bonanni, G. Rozza, A LSTM-enhanced surrogate model to simulate the dynamics of particle-laden fluid systems, (2024). http://arxiv.org/abs/2403.14283 (accessed August 12, 2024).